\documentclass[epj]{svjour}

\usepackage{amsmath,amssymb,amsfonts}	
\usepackage{graphicx}
\usepackage{xspace}
\usepackage{booktabs}
\usepackage{makecell}
\usepackage{slashed}

\usepackage[colorlinks=true,pdfstartview=FitV, linkcolor=blue, citecolor=blue,urlcolor=blue]{hyperref} 
\usepackage[sort&compress,numbers,colon,merge]{natbib}

\begin{document}

\title{Towards the optimal beam dump experiment to search for feebly interacting particles}
 \author{Kyrylo~Bondarenko\inst{1,2,3} \and Alexey~Boyarsky\inst{4} \and Richard Jacobsson\inst{5} \and Oleksii Mikulenko\inst{4} \and Maksym Ovchynnikov\inst{6,4}}
  \institute{IFPU, Institute for Fundamental Physics of the Universe, via Beirut 2, I-34014 Trieste, Italy
  \and
  SISSA, via Bonomea 265, I-34132 Trieste, Italy
  \and
  INFN, Sezione di Trieste, SISSA, Via Bonomea 265, 34136, Trieste, Italy
  \and 
  Instituut-Lorentz, Leiden University, Niels Bohrweg 2, 2333 CA Leiden, The Netherlands
  \and 
  European Organization for Nuclear Research (CERN), Geneva, Switzerland
  \and 
  Institut für Astroteilchen Physik, Karlsruher Institut für Technologie (KIT), Hermann-von-Helmholtz-Platz 1, 76344 Eggenstein-Leopoldshafen, Germany
  }
  \abstract{Future searches for new physics beyond the Standard Model are without doubt in need of a diverse approach and experiments with complementary sensitivities to different types of classes of models. One of the directions that should be explored is feebly interacting particles (FIPs) with masses below the electroweak scale. The interest in FIPs has significantly increased in the last ten years. Searches for FIPs at colliders have intrinsic limitations in the region they may probe, significantly restricting exploration of the mass range $m_{\text{FIP}} < 5-10$\,GeV/c$^2$.
Beam dump-like experiments, characterized by the possibility of extremely high luminosity at relatively high energies and the effective coverage of the production and decay acceptance, are the perfect option to generically explore the ``coupling frontier'' of the light FIPs. Several proposals for beam-dump detectors are currently being considered by CERN for implementation at the SPS ECN3 beam facility. In this paper, we analyse in depth how the characteristic geometric parameters of a beam dump experiment influence the signal yield. We apply an inclusive approach by considering the phenomenology of different types of FIPs. From the various production modes and kinematics, we demonstrate that the optimal layout that maximises the production and decay acceptance consists of a detector located on the beam-axis, at the shortest possible distance from the target defined by the systems required to suppress the beam-induced backgrounds.}

\maketitle

\section{Introduction}
\label{sec:introduction}

Despite the success of the Standard Model (SM) of particle physics, evidence for the existence of new physics beyond the Standard Model is already well established because the origin of neutrino oscillations, dark matter, and the baryon asymmetry of the Universe is not known. However, we have no solid predictions of where to search for it. New particles capable of resolving these problems can have masses from sub-eV to Planck scale and coupling constants with SM particles ranging many orders of magnitude. At this crossroad point of particle physics, it is essential to use efficiently available or planned experimental facilities to push forward different frontiers of physics, probing whole classes of models simultaneously.

If the mass of a new particle is below the EW scale, it may be produced at accelerators not only as a resonance but also in decays of SM particles, such as heavy bosons $W,Z,h$, as well as mesons $\pi,D,B$. This makes this range of masses of new particles especially interesting from an experimental point of view. 
In this mass range, new particles would have escaped detection, not because of the limit on available accelerator energy, but because their creation is extremely rare. Numerous searches at past experiments as well as at the LHC constrain large values of coupling constants, which is why new particles of this type are often called feebly-interacting particles, or FIPs (see e.g.~\cite{Beacham:2019nyx,Agrawal:2021dbo}).

FIPs can play a direct role in the beyond SM phenomena, like e.g. heavy neutral leptons or HNLs in the sub-EW mass range explain neutrino masses via sea-saw mechanism and matter-antimatter asymmetry via their out-of-equilibrium kinetics in the early Universe at a temperature above 100 GeV. They can also be a "portal" that connects the SM sector with a Dark Sector, i.e. the case in which the Dark Sector particles only interact with ordinary matter via the FIP mediator (see e.g.~\cite{Alekhin:2015byh}). FIPs, with their tiny coupling constants, form a "coupling frontier" of particle physics, and are part of a whole class of SM extensions.

If there is no particular reason (e.g. symmetry) for making these new particles stable, their lifetime scales with the coupling constant $g$ and mass $m_{\text{FIP}}$ as $\tau_{\text{FIP}} \propto g^{-2}m_{\text{FIP}}^{-\alpha}$ where $\alpha=1-5$~\cite{Alekhin:2015byh}. Depending on the lifetime, different search strategies should be used to probe the FIP parameter space efficiently. In particular, particles with lifetimes $c\tau_{\text{FIP}} \gtrsim 1\text{ mm}$ can be searched for via displaced-vertices schemes at the LHC and future colliders~\cite{CMS:2022fut,Boiarska:2019jcw,Blondel:2022qqo,Abdullahi:2022jlv}. The limitation on $c\tau_{\text{FIP}}$ from below comes from numerous backgrounds caused by events with SM particles, which occur at small displacements. The limitation from above is FIP-dependent. It is caused by two factors. First, the collider detectors have short decay volumes of the order of $\mathcal{O}(1\text{ m})$, typically defined by the dimensions of their inner trackers~\cite{CMS:2022fut}. As a result, long-lived FIPs mainly decay well beyond the fiducial volume. Second, to reduce SM backgrounds, one must impose severe selection criteria on candidate events with FIPs, such as kinematic properties and specific final states, resulting in low signal efficiency. For instance, a typical selection efficiency of recent searches for HNLs at CMS~\cite{CMS:2022fut} was of the order of 1\%. Therefore, even if a long-lived FIP decays inside the tracker, the event will likely be outside the selection acceptance.

\begin{figure*}[t]
    \centering
    \includegraphics[width=0.45\textwidth]{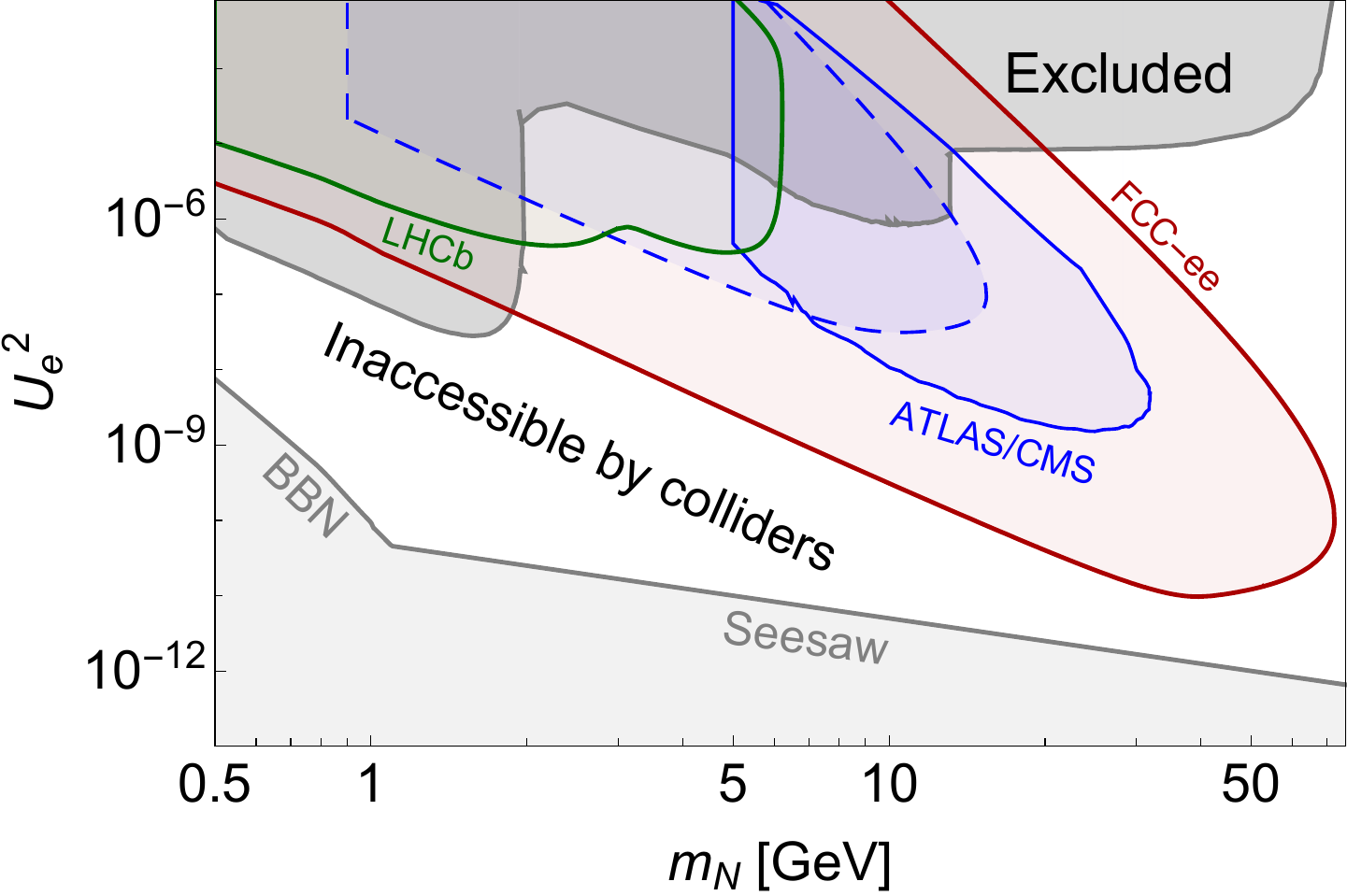}~\includegraphics[width=0.45\textwidth]{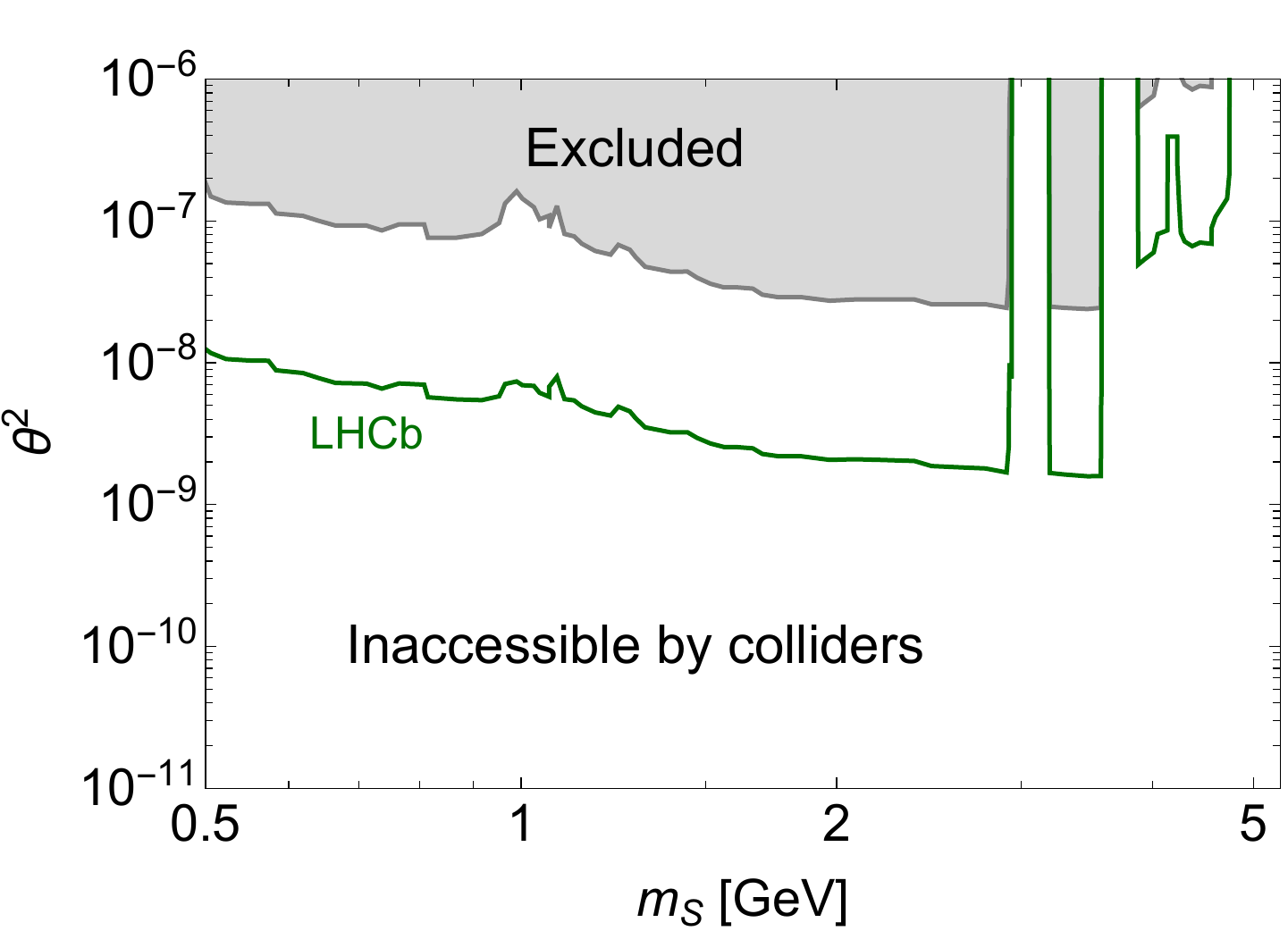}
    \caption{The potential of future collider searches to probe the parameter space of feebly-interacting particles in the plane FIP mass - FIP coupling to SM particles. The figures demonstrate that colliders cannot efficiently explore the parameter space of FIPs with mass of the order of GeV. \textbf{Left panel:} parameter space of heavy neutral leptons (HNLs) that mix with electron neutrinos. The lower bound (seesaw) is defined by their ability to generate masses for active neutrinos~\cite{Asaka:2005an}. LHC in high luminosity phase~\cite{Boiarska:2019jcw,Abdullahi:2022jlv} and lepton colliders~\cite{Blondel:2022qqo} are mainly sensitive to short-lived HNLs, with the typical lifetimes $c\tau_{N}\lesssim \mathcal{O}(100\text{ m})$, see text for details. The scaling of the HNL lifetime with the mass is $\tau_{N}\propto m_{N}^{-5}U_{e}^{-2}$. As a result, colliders have poor sensitivity to HNL masses $m_{N}\lesssim 10\text{ GeV}$. \textbf{Right panel}: dark scalars mixing with Higgs bosons. Given the strict event selection to cope with the backgrounds and the available triggers, the event rate with displaced vertices at colliders is insufficient to provide a competitive sensitivity. Instead, scalars may be searched for with prompt events at LHCb~\cite{LHCb:2016awg}, even though only relatively large couplings are within reach. The parameter space of these examples and other feebly-interacting particles in the GeV range can instead be more efficiently explored in the coming years with beam dump experiments.}
    \label{fig:light-FIPs-colliders}
\end{figure*}

Typically, the FIP lifetime and its production rate are controlled by the same coupling, i.e. a significantly long FIP lifetime also means a small production rate.\footnote{A counter-example is the model of dark scalars with the mixing and quartic coupling to the Higgs bosons, $\mathcal{L} = \theta m_{h}^{2}Sh + \alpha hS^{2}$. Depending on the value of $\alpha$, the production may be dominated by the Higgs boson decay $h\to SS$, while the decay is mediated by $\theta$.} 
Together with typically low signal efficiencies coming from the trigger and the event-versus-background selection, it is  possible that the FIP production rate within the acceptance of the searches is insufficient to provide any sensitivity. This is the case of dark scalars with the coupling through mixing, dark photons, and axion-like particles~\cite{Beacham:2019nyx}. Even for FIPs where the production is in principle sufficient (e.g., heavy neutral leptons~\cite{Boiarska:2019jcw}), future collider searches have a limited potential to probe the parameter space of GeV-scale FIPs. This is explained by the behavior of the FIP decay length $c\tau_{\text{FIP}}\gamma_{\text{FIP}} \propto g^{-2}m_{\text{FIP}}^{-\alpha-1}$ -- for a given coupling $g$, the lifetime rapidly increases with decreasing $m_{\text{FIP}}$, in other words, quickly reducing the FIP decay probability within the fiducial volume and hence the sensitivity. We illustrate these points in Fig.~\ref{fig:light-FIPs-colliders}, where we show the parameter space of heavy neutral leptons (HNLs) and dark scalars mixing with Higgs bosons. Future development of new trackers~\cite{Alimena:2021mdu,LHCP:2022} at ATLAS, CMS, and LHCb may improve the LHC reach for light FIPs, but the remaining explorable region in the parameter space will still be large (see Sec.~\ref{sec:conclusions}). Altogether, this suggests that we need a special experiment to search for FIPs in the GeV range.

In this paper, we argue that the most suitable experimental setup to search for FIPs with mass below 5\,GeV is a beam dump experiment, where an extracted proton beam hits a dense target, and where the search is performed in a displaced decay volume. Although having much lower centre-of-mass energy of collisions than at colliders, beam dumps can deliver extremely high luminosity by operating with a more intensive proton beam combined with a high-A/Z target. This means in particular that they are capable of delivering a large number of mesons within a relatively small forward solid angle, in particular $B,D$, that may further decay into FIPs in the mass range of interest. There is no limitation on the decay volume length. It can easily be several tens of metres to cover a much larger lifetime acceptance than at collider detectors. Finally, backgrounds can be significantly reduced by placing the decay volume behind a chain of components designed to suppress beam-induced particle backgrounds, such as a hadron absorber and a muon deflector, considerably reducing the need to impose strict signal selection criteria. The best placement of such an experiment is the SPS accelerator at CERN, operating with a high-intensity proton beam of energy $E_{p} = 400\text{ GeV}$.

Several proposals of beam dump experiments at the SPS have been made~\cite{Aberle:2839677,CortinaGil:2839661,Alviggi:2839484}. They differ in geometric parameters, including the placement with respect to the beam axis, the decay volume size, and the detector angular coverage, both in terms of production acceptance and decay acceptance. The experiments also differ in the choice of material for the proton target. Maximum production of FIPs, and simultaneously maximum suppression of background from pion and kaon decays to muons and neutrinos, are achieved with a target of the highest possible atomic mass and atomic number, as well as minimized internal cooling for density. However, to discuss the optimal experimental layout, we assume in this paper the same target material for all experiments.

\begin{table}[!h]
    \centering
    \begin{tabular}{|c|c|c|c|c|c|c|}
       \hline $l_{\text{min}}$& $S_{\text{det}}$ & $l_{\text{fid}}$ & $l_{\text{det}}$ & $r_{\text{displ}}$  \\ \hline
    38 m& $4\times 6\text{ m}^{2}$ & 50 m & 15 m & 0 m \\ \hline
    \end{tabular}
    \caption{Parameters of the hypothetical experiment used as a reference experiment in our estimates: the longitudinal distance from the target to the beginning of the decay volume; the transverse dimensions of the decay volume and the detector, the longitudinal length of the decay volume; the longitudinal length of the detector; the distance from the centre of the detector in the transverse plane to the beamline. Here and below, we assume that the decay volume is oriented parallel to the beamline, which is motivated by the typical constraints from available space and infrastructure.}
    \label{tab:hypothetical-experiment-parameters}
\end{table}

In this paper, we perform an analysis of the sensitivity to FIPs with respect to the geometric parameters of a beam dump experiment, in a maximally model-independent way, in order to find the optimal configuration. We start with an on-axis experiment specified in Table~\ref{tab:hypothetical-experiment-parameters}. Its close analog is the SHiP experiment~\cite{SHiP:2015vad,Aberle:2839677}. We then study how the FIP sensitivity is affected by changing the parameters from the table. To be model-independent, we consider a few FIP models covering a wide class of production mechanisms and decay modes.

Our main results are summarized in Figures~\ref{fig:on-axis},~\ref{fig:off-axis},~\ref{fig:upper-bound},~\ref{fig:summary}. We demonstrate that the setup in Table~\ref{tab:hypothetical-experiment-parameters} is optimal for searching for FIPs independently of their lifetime, being also compatible with the absence of backgrounds. One of the main reasons for providing the largest signal acceptance is the on-axis placement. The off-axis location leads to a substantial loss of acceptance, and significantly worsens the ability to reconstruct properties of FIPs such as mass, spin, and decay modes, see Figure~\ref{fig:summary}. 

The paper is organized as follows. In Sec.~\ref{sec:qualitative-analysis}, we start from the expression for the number of events and discuss the FIP phenomenology (subsection~\ref{sec:fip-phenomenology}), describing, in particular, their production and decay modes. In Sec.~\ref{sec:analysis}, we analyze how the number of events at the lower bound of the sensitivity varies with the experimental configuration, considering separately its placements on-axis (Sec.~\ref{sec:on-axis}) and off-axis (Sec.~\ref{sec:off-axis}) relatively to the beamline. In Sec.~\ref{sec:upper-bound}, we study the impact of the configuration on the potential to probe FIPs at the upper bound of the sensitivity. In Sec.~\ref{sec:comparison-proposals}, we apply our findings to compare the sensitivity of the experiments proposed at SPS, SHiP, SHADOWS, and HIKE. Finally, in Sec.~\ref{sec:conclusions}, we make conclusions. Appendices contain all the relevant technical information on the phenomenology of FIPs and the calculations.

\section{Number of signal events}
\label{sec:qualitative-analysis}

Let us start with the expression for the number of events in the regime of large lifetimes, where the typical decay length of FIPs, $c\tau_{\text{FIP}}\langle\gamma_{\text{FIP}}\rangle$, is much larger than the characteristic scale of the experiments $\simeq 100\text{ m}$ (the opposite cases of short lifetimes $c\tau_{\text{FIP}}\langle\gamma_{\text{FIP}}\rangle\lesssim l_{\text{min}}$ is discussed in Sec.~\ref{sec:upper-bound}). In this case, the number of events may be represented in the following schematic form (see Appendix~\ref{app:numerics}):
\begin{equation}
    N_{\text{events}} \approx N_{X,\text{prod}} \times \frac{l_{\text{fid}}\langle p_{\text{FIP}}^{-1}\rangle}{c\tau_{\text{FIP}}m_{\text{FIP}}}\times \epsilon_{\text{geom}} \times \epsilon_{\text{rec}} \times \text{Br}_{\text{vis}}
    \label{eq:Nevents-approx}
\end{equation}
Here, $N_{X,\text{prod}} = N_{\text{PoT}}\times \text{Br}(pp\to X)$ is the total number of the FIPs produced in the collisions, with $N_{\text{PoT}}$ being the number of protons on target, and $\text{Br}(pp\to X)$ the probability of producing of FIPs by any mechanism; the second factor is the decay probability in the regime of large lifetimes, with $\langle p_{\text{FIP}}^{-1}\rangle$ being the mean inverse momentum for the FIP at the experiment; $\epsilon_{\text{geom}}$ is the overall geometric acceptance, following from geometric limitations of the decay volume and detectors (discussed in details below); $\text{Br}_{\text{vis}}$ is the branching ratio of the decay of FIPs into states visible at the given experiment; finally, $\epsilon_{\text{rec}}$ is the total reconstruction efficiency -- the fraction of events within the geometric acceptance that may be reconstructed. 

Independently of the configuration, we will assume $\epsilon_{\text{rec}}= 1$, and that $\text{Br}_{\text{vis}}$ includes all decays with at least two electrically charged particles or two photons.

Considering two experiments located at the same facility (such that the beam energy and target configuration are the same), for the ratio of the number of events~\eqref{eq:Nevents-approx} one has
\begin{equation}
    \frac{N_{\text{events,1}}}{N_{\text{events,2}}} \approx \frac{l_{\text{fid,1}}}{l_{\text{fid,2}}}\times \frac{\epsilon_{\text{geom,1}}}{\epsilon_{\text{geom,2}}} \times \frac{\langle E_{\text{FIP}}^{-1}\rangle_{1}}{\langle E_{\text{FIP}}^{-1}\rangle_{2}}
\end{equation}

Therefore, to compare the lower bounds of sensitivity of two experiments, we need to understand the behavior of $\epsilon_{\text{geom}}$ and $\langle E_{\text{FIP}}^{-1}\rangle$. 

\subsection{Geometric acceptance}
\label{sec:geom-acceptance}
\begin{figure*}[!t]
    \centering
    \includegraphics[width=0.7\textwidth]{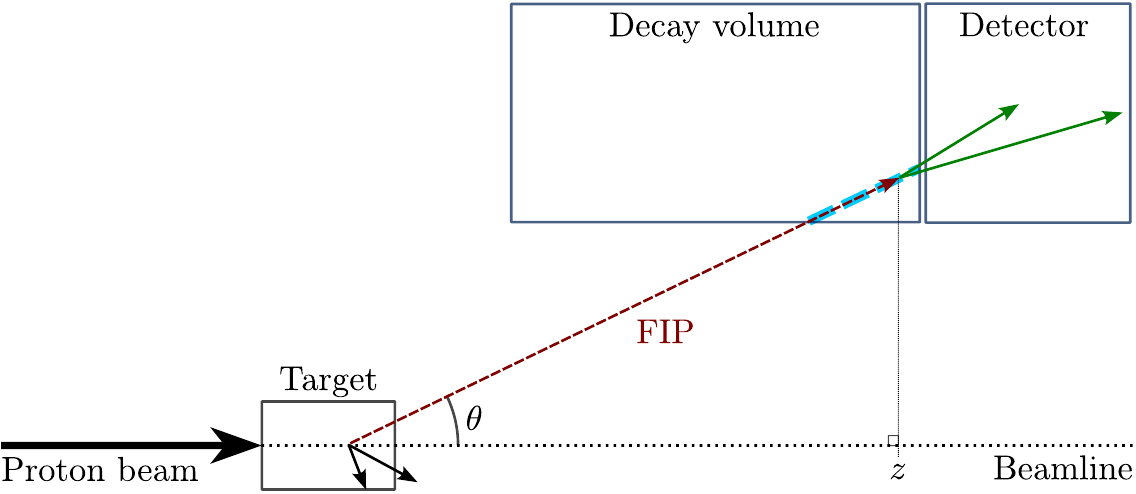}
    \caption{Illustration of the impact of different contributions to the geometric acceptance defined by Eq.~\eqref{eq:acceptance-qualitative}. First, the FIPs produced by collisions of the proton beam with the fixed target must point to the detector (the red arrow). The fraction of such events is given by $\epsilon_{\text{FIP}}$. The effective length inside the decay volume passed by decaying FIPs (the dashed blue line) may differ significantly from the nominal decay volume length $l_{\text{fid}}$. This results in the factor $l_{\text{fid,eff}}/l_{\text{fid}}$. Finally, the decay products of FIPs (the green arrows) also have to point to the detector, which is incorporated by $\epsilon_{\text{dec}}$.}
    \label{fig:acceptance-illustration}
\end{figure*}

Schematically, the geometric acceptance is given by
\begin{equation}
    \epsilon_{\text{geom}} \simeq \epsilon_{\text{FIP}}\times l_{\text{fid}}^{\text{eff}}/l_{\text{fid}} \times \epsilon_{\text{dec}},
    \label{eq:acceptance-qualitative}
\end{equation}
see Fig.~\ref{fig:acceptance-illustration}.

The first factor is the FIP acceptance, i.e., the fraction of FIPs with trajectories pointing to the cross-section of the end of the detector.\footnote{Naively, the mentioned definition of the FIP acceptance looks too restrictive, as it does not take into account the FIPs decaying inside the decay volume but not pointing to the detector. However, because of the 4-momentum conservation, FIPs that do not point to the detector typically cannot decay into particles pointing to the detector. Therefore, instead of considering FIPs decaying in any direction inside the decay volume, we considered only the FIPs that already point to the detector.} $l_{\text{fid}}^{\text{eff}}$ is effective fiducial volume length -- the mean length inside the decay volume passed by FIPs pointing to the end of the detector. If the considered experiment is located on-axis, then $l_{\text{fid}}^{\text{eff}} \approx l_{\text{fid}}$. However, in case of the off-axis placement, parallel to the beamline, $l_{\text{fid}}^{\text{eff}}$ gets effectively reduced. Finally, the last factor is the decay product acceptance, i.e., the fraction of decays of FIPs within $\epsilon_{\text{FIP}}$ with at least two of their decay products pointing to the end of the detector. Roughly, the decay products of FIPs with the gamma factor $\gamma_{\text{FIP}}$ have the opening angle\footnote{For simplicity, we considered here 2-body decays; however, for 3-body decays, the situation is qualitatively similar.}
\begin{equation}
\Delta\theta_{\text{dec}} \sim 2/\gamma_{\text{FIP}}
\label{eq:decay-angle}
\end{equation}
If this angle becomes comparable with the angle covered by the detector as seen from the FIP's decay point, then $\epsilon_{\text{dec}}$ would significantly reduce the event rate. 

\section{Kinematic distributions of FIPs}
\label{sec:fip-phenomenology}

\begin{figure*}[t]
    \centering
  \includegraphics[width=0.9\textwidth]{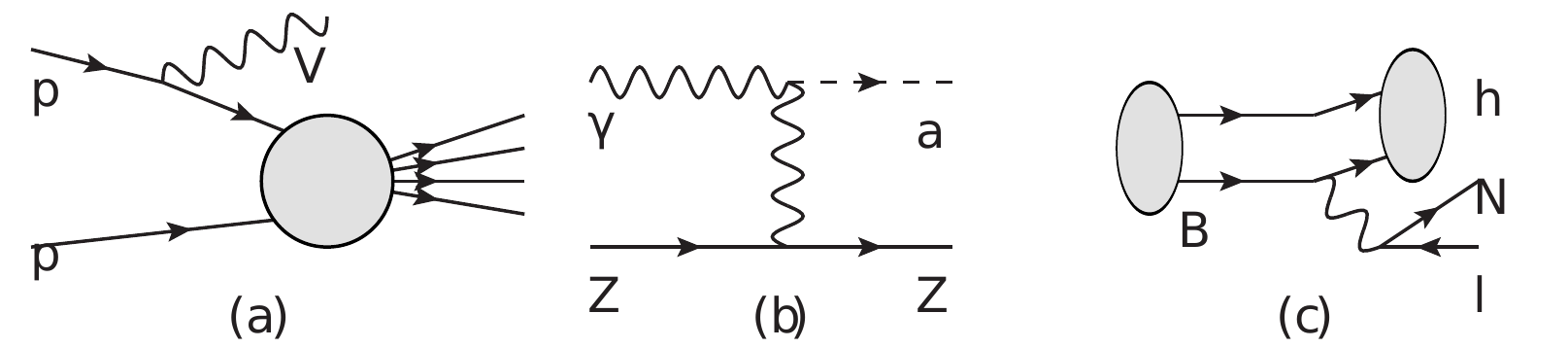}
    \caption{Examples of production processes for various FIPs: (a) proton bremsstrahlung (dark photon $V$), (b) coherent scattering off nuclei (ALP with the photon coupling $a$), (c) decays of $B$ mesons (HNLs $N$, dark scalars).}
    \label{fig:production-channels}
\end{figure*}

To understand the behavior of $\epsilon_{\text{FIP}}$ and $\epsilon_{\text{dec}}$, we need to study in a model-independent way how FIPs may be produced in proton-target collisions and how they decay. To this extent, we consider different types of FIPs: dark photons, dark scalars, HNLs, and ALPs with photon coupling~\cite{Beacham:2019nyx}. By considering all of them, we can perform the analysis in a maximally model-independent fashion. The dominant production mechanisms and decay modes are shown in Fig.~\ref{fig:production-channels} and in Table~\ref{tab:fip-model-channels}. The production channels are decays of mesons (light unflavoured mesons $\pi,\eta,\eta',\rho^{0}$, as well as heavy flavored mesons $B,D$), or the direct production through proton bremsstrahlung, Drell-Yan process, coherent proton-nucleus and photon-nucleus scattering. We generate the distribution of the light mesons at the SPS using the approach of~\cite{Dobrich:2019dxc} (see also~\cite{Jerhot:2022chi}), and use the distribution of $B,D$ mesons from~\cite{CERN-SHiP-NOTE-2015-009}. We follow the description of the bremsstrahlung process from~\cite{SHiP:2020vbd} and the coherent scattering from~\cite{Dobrich:2019dxc}. For the details on the derivation of the distributions of FIPs from these production channels, see Appendix~\ref{app:numerics}. 

{\small

\begin{table*}[!t]
    \centering
    \begin{tabular}{|c|c|c|c|c|c|}
      \hline FIP & Prod. modes & Decay modes \\ 
      \hline \textbf{DP $V$} & $\begin{cases}\pi^{0}/\eta\to V, \ m_{V}<m_{\eta} \\ \text{Brem/DIS}, \ m_{V}>m_{\eta}\end{cases}$ & \makecell{$V\to ll$ \\ $V\to 2\pi, 3\pi, KK, m_{V}\lesssim 1\text{ GeV}$ \\ $V\to qq, m_{V}\gtrsim 1\text{ GeV}$} \\ \hline  \textbf{ALP$_{\gamma}$ $a$} & \makecell{$\gamma+Z\to a+Z$\\ $p+Z\to p+Z+a$} & $a\to \gamma\gamma$\\  \hline \textbf{Scalar $S$} & $\begin{cases} K\to S+\pi, \ m_{S}<m_{K}-m_{\pi} \\ B\to S+X, \ m_{S}>m_{K}+m_{\pi}\end{cases}$ & \makecell{$S\to ll$ \\ $S < \pi\pi/KK, m_{S}<2\text{ GeV}$ \\ $S\to qq, m_{S}>2\text{ GeV}$} \\ \hline \textbf{HNL $N$} & $\begin{cases} K\to N+X,\ m_{N}<m_{K},\\ D\to N+X, m_{N}<m_{D_{s}}, \\ B\to N+X, \ m_{N}>m_{D_{s}}\end{cases}$  & \makecell{$N\to ll\nu$\\  $N\to \pi^{0}\nu,\eta \nu, \pi l, m_{N}\lesssim 1\text{ GeV}$\\ $N\to qq\nu, qql,m_{N}\gtrsim 1\text{ GeV}$} \\ \hline 
    \end{tabular}
    \caption{Dominant production and decay modes of GeV-scale FIPs in proton-proton collisions at the SPS. We consider the dark photon, the ALP with the photon coupling, the Higgs-like scalar with mixing coupling, and HNLs with the dominant coupling to electron neutrinos. For more details, see~\cite{Ovchynnikov:2023cry}.}
    \label{tab:fip-model-channels}
\end{table*}}

\begin{figure*}[t]
    \centering
    \includegraphics[width=0.45\textwidth]{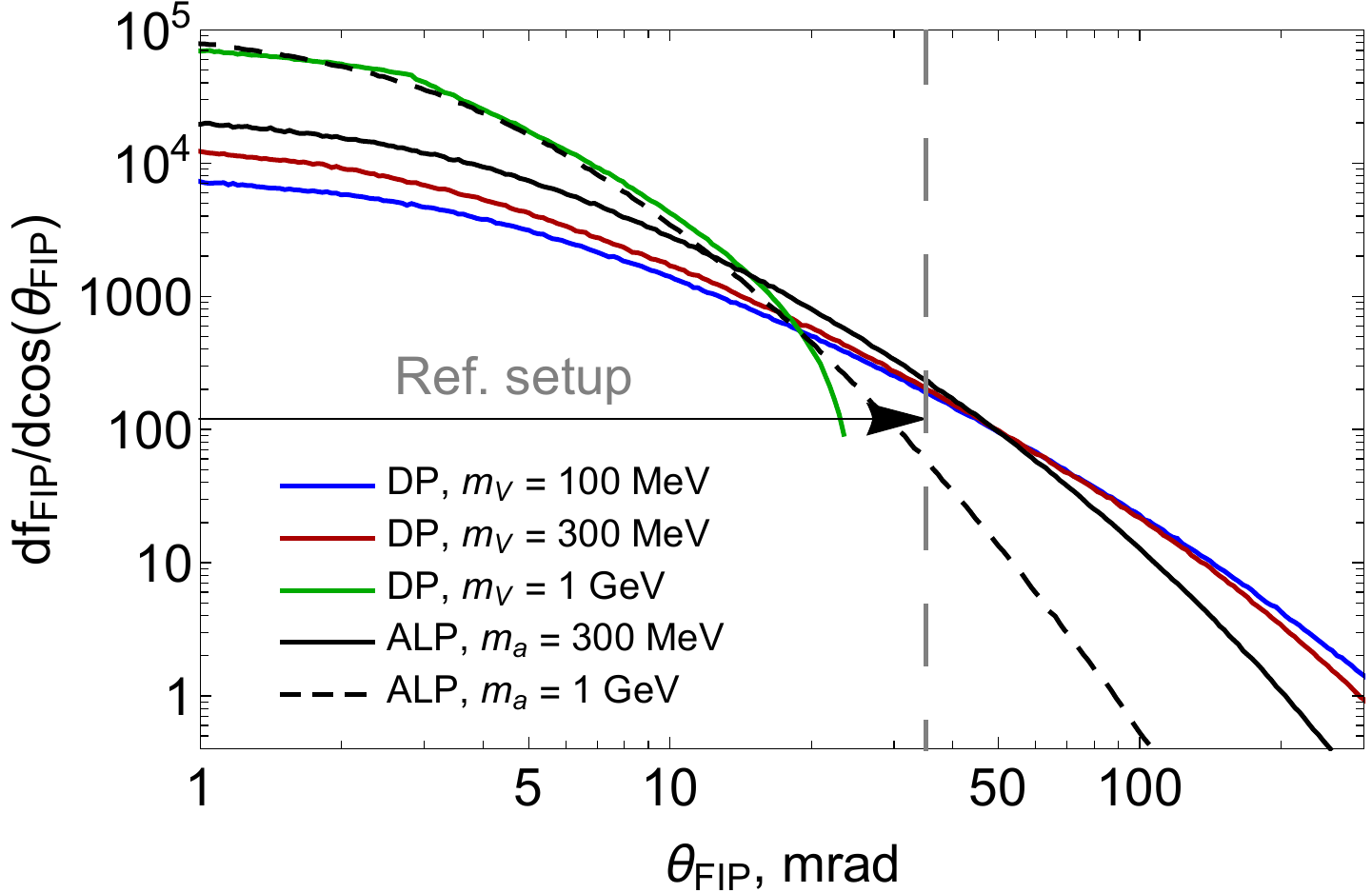}~\includegraphics[width=0.45\textwidth]{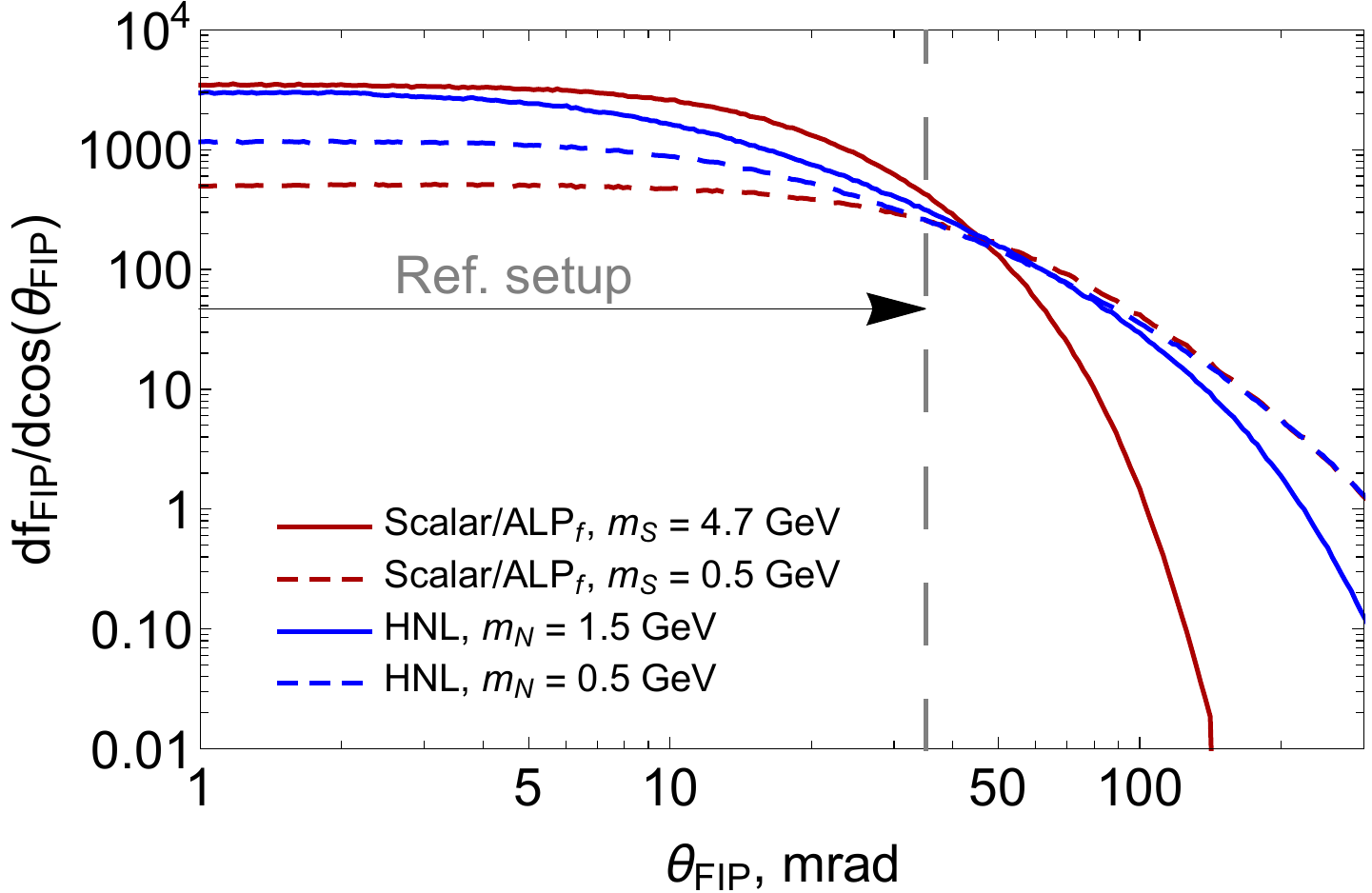}\\ \includegraphics[width=0.45\textwidth]{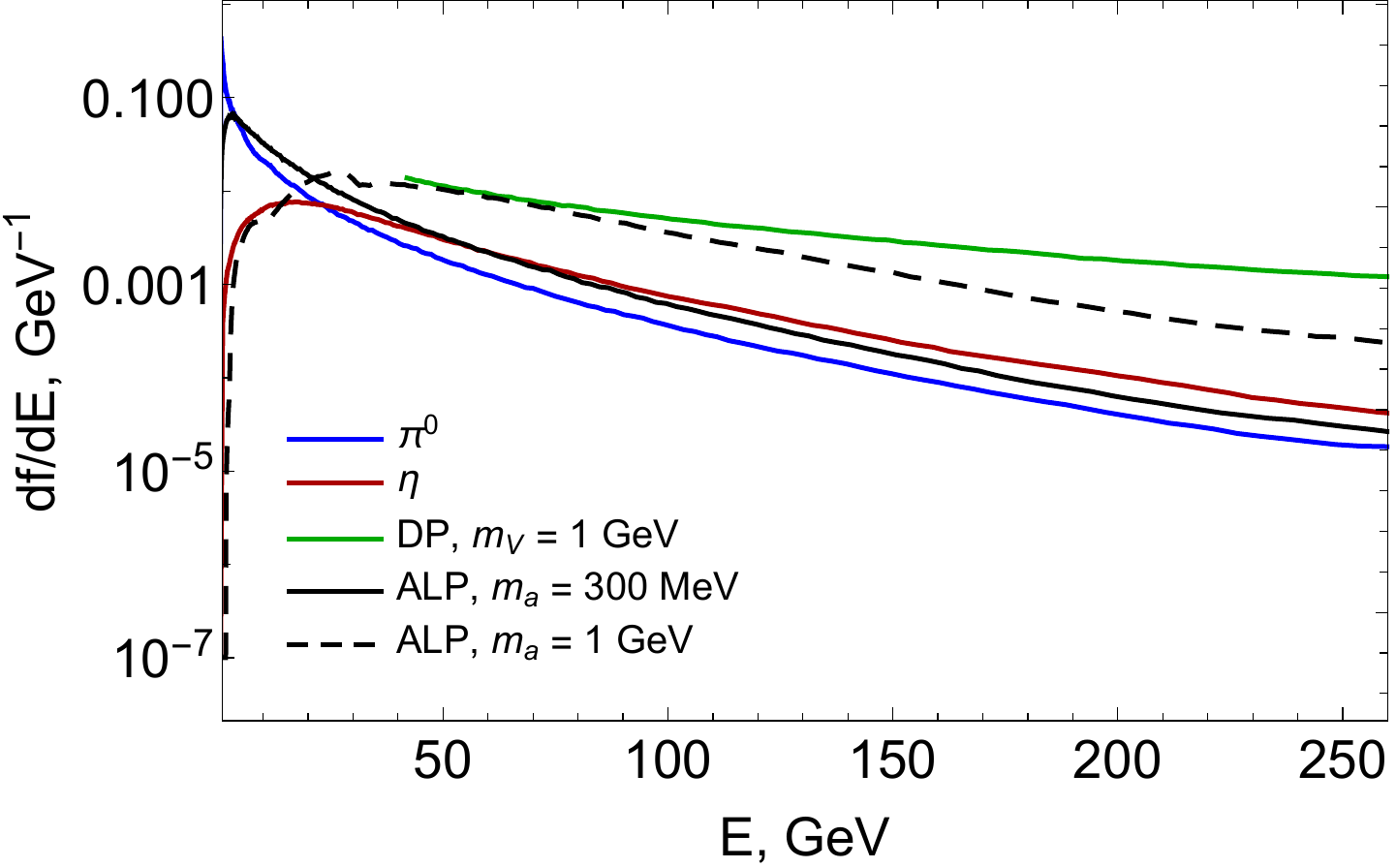}~\includegraphics[width=0.45\textwidth]{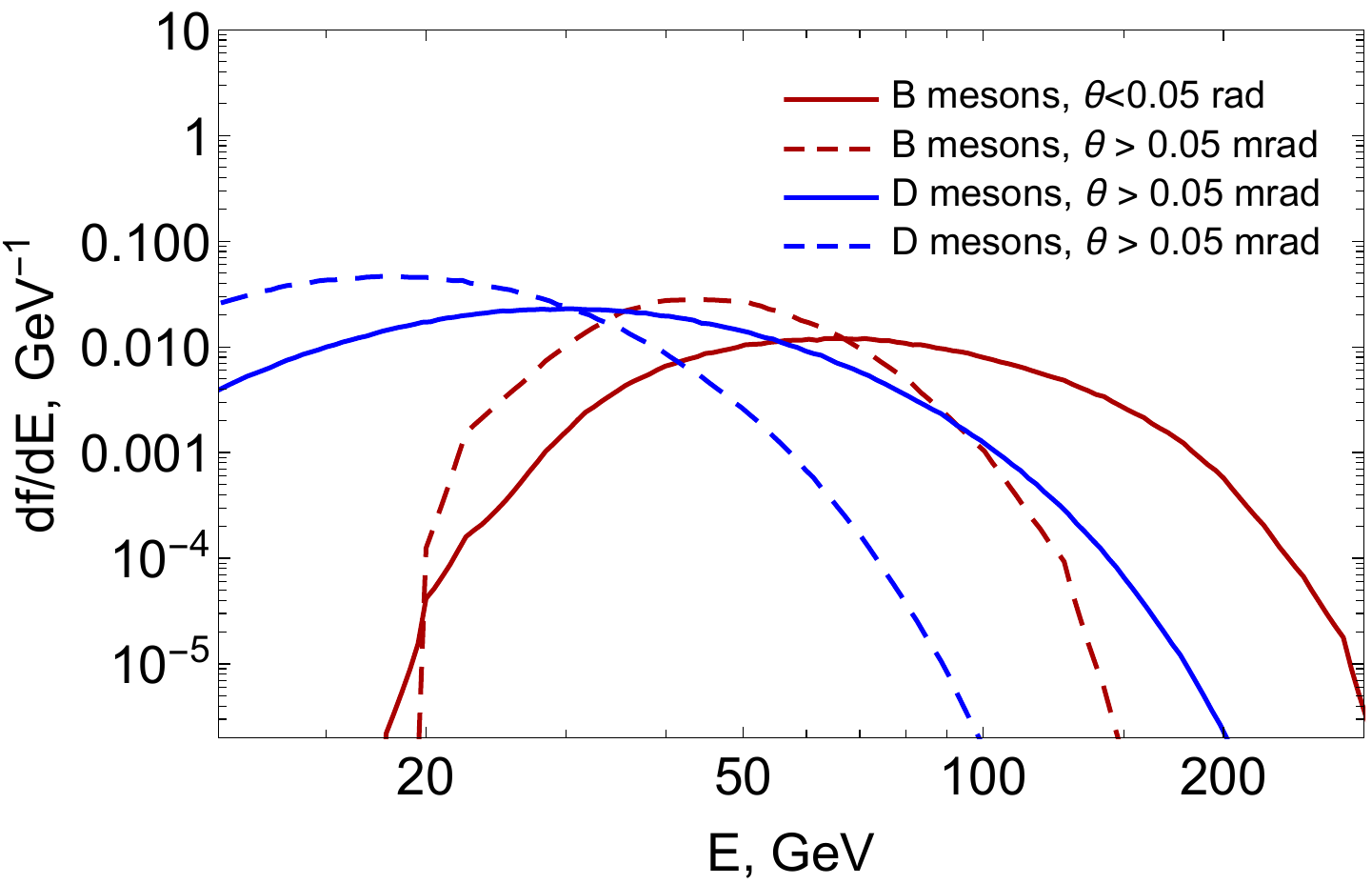}
    \caption{Kinematics of FIPs produced in proton-target collisions at the SPS. Molybdenum target is considered. \textbf{Top panels}: solid angle distributions $df_{\text{FIP}}/d\Omega_{\text{FIP}}\sim df_{\text{FIP}}/d\cos(\theta_{\text{FIP}})$ of various FIPs. Different masses are considered, corresponding to different production channels (Table~\ref{tab:fip-model-channels}). Note that the distribution of heavy HNLs with $m_{N}>3\text{ GeV}$ is very similar to the distribution of scalars, because of the same mother particle and decay kinematics. The polar angle coverage of the detector of the reference setup~\ref{tab:hypothetical-experiment-parameters} is indicated with arrows and the vertical dashed line. \textbf{Bottom panels}: energy spectra of the mesons producing FIPs, dark photons produced by the proton bremsstrahlung, and ALPs with photon coupling. For the case of heavy mesons $B,D$, the distribution is shown assuming two different angular coverage: ``on-axis'' $\theta < 0.05\text{ rad}$, and ``off-axis'', $\theta >0.05\text{ rad}$, to demonstrate how the spectrum gets softer off-axis. See text and Ref.~\cite{Ovchynnikov:2023cry} for details.}
    \label{fig:FIPs-distributions}
\end{figure*}

The solid angle distributions 
\begin{equation}
df_{\text{FIP}}/d\Omega_{\text{FIP}}\sim df_{\text{FIP}}/d\cos(\theta_{\text{FIP}}))\end{equation} 
of the FIPs produced by these mechanisms, as well as their energy distributions are shown in Fig.~\ref{fig:FIPs-distributions}. Because of the kinematics of the collisions with a fixed target, the bulk of the distributions of FIPs are contained within a relatively small forward solid angle around the beam axis, being flat up to and quickly dropping at large angles $\theta>\theta_{\text{flat}}$. The direct production processes are characterized by very small typical transverse momentum compared to the momentum of the incoming proton. In the case of Drell-Yan and proton bremsstrahlung, it is of the order of $m_{p} \sim 1\text{ GeV}$~\cite{Gorbunov:2014wqa}. Given very large typical energies of the FIPs produced by these mechanisms, $E_{\text{FIP}}\sim 100-200\text{ GeV}$, the angle is $\theta_{\text{flat}}\simeq 10\text{ mrad}$. For the Primakov production off nuclei and nucleons, $\theta_{\text{flat}}$ is determined by typical transverse momenta carried by virtual photons, being $\theta_{\text{flat}}\simeq m_{p}/E_{\text{beam}} = 2.5\cdot 10^{-3}\text{ rad}$~\cite{Dobrich:2015jyk}.

Light mesons produced in collisions have characteristic $p_{T}$ of the order of $\Lambda_{\text{QCD}}$, and relatively small mean energy of order of~$\langle E\rangle \simeq 20-30\text{ GeV}$, which leads to $\theta_{\text{flat}}$ being of the order of a few tenths of mrad~\cite{Dobrich:2019dxc}.

Heavy mesons $D,B$ have large $p_{T}$ of the order of their mass, but, at the same time, much larger characteristic energies. As a result, their distributions start dropping at even smaller angles $\theta_{\text{flat}}\lesssim 10\text{ mrad}$. For lighter masses, the FIPs produced by decays of $B,D$ may have broader angular distribution. The reason is an additional transverse momentum of the order of the energy of the FIP at the rest frame of the decaying meson, which may be as large as the meson mass for FIPs with $m_{\text{FIP}}\ll m_{\text{meson}}$.

The main uncertainty for the FIP production (which translates to the main uncertainty in the signal rate estimates) comes from the thickness of the target, which gives rise to a cascade production of the SM particles that may potentially decay into FIPs. The cascade population affects both the total yield of the particles and the shape of the angle-energy distribution: the angular distribution of these extra particles is broader, while the energy distribution is shifted to lower energies.

While for the thin target and SPS beam energies the fluxes of mesons have been measured~\cite{Adamovich:1992cv,Aguilar-Benitez:1991hzq}, observing the cascade flux in the case of thick targets is more complicated (as the mesons or their decay products may be absorbed inside the target) and requires a dedicated experiment. Therefore, we may only estimate it via simulations. Such estimates have already been performed by the SHiP collaboration~\cite{SHiP:2020vbd,CERN-SHiP-NOTE-2015-009}. They suggest that for heavy flavor production, the cascade enhancement gives a factor of 2 to the total meson yield and is important for both on-axis and off-axis experiments (see also~\cite{Dobrich:2018jyi}); we account for these cascade mesons in our calculations. For lighter mesons, the enhancement may be a factor of 10 or larger~\cite{SHiP:2020vbd}. However, these mesons are very soft, with energies of order of 1 GeV, and decay into a very low-energy FIP produced in a broad range of polar angles. Even if such FIPs would decay inside the decay volume, the products would not be properly reconstructed because of low energies.\footnote{For instance, selection criterion at SHiP for the decay products is $E>1\text{ GeV}$.} As a result, only a tiny fraction of such cascade mesons contributes to the signal yield. Therefore, we may safely neglect them.

The decay modes of FIPs may differ by the number of the decay products and their phase space if the number of products is fixed, which comes from the ratio of the FIP-to-decay product masses and the matrix element of the decay. The main decay modes of the FIPs are provided in Table~\ref{tab:fip-model-channels}.

Another uncertainty in calculating the number of events may come from theoretical uncertainties in the description of the phenomenology of FIPs. Namely, for the particles such as Higgs-like scalars, various ways to describe hadronic decays may change the decay width by 1-2 orders of magnitude (see, e.g.,~\cite{Boiarska:2019jym,Gorbunov:2023lga} and references therein), which affects both the lower and upper bounds of the sensitivity. While it may be important for the full sensitivity region, it is not relevant when computing the ratio of the number of events in the regime of long lifetimes, as we do in this paper -- just the parameter space gets shifted.

\section{Effect on the number of signal events from varying parameters of the experiment}
\label{sec:analysis}

In this section, we study the impact of varying the geometric parameters of the experiment on the number of events. To illustrate the effect of the decay products acceptance and the FIP angular distribution, we consider the following FIPs and properties:

\begin{itemize}
    \item[--] HNLs with masses $0.5\text{ GeV}$, $1.5\text{ GeV}$ produced by decays of $D$ mesons, and $4\text{ GeV}$ produced by decays of $B$.
    \item[--] Dark scalars with masses $0.5\text{ GeV}$ and $4\text{ GeV}$ produced by decays of $B$.
    \item[--] Dark photons with masses $m_{V} = 300\text{ MeV}$ produced by decays of $\eta$ and $m_{V} = 1\text{ GeV}$ by proton bremsstrahlung.
    \item[--] ALPs with photon coupling with masses $m_{a} = 300\text{ MeV}$ and 1 GeV.
\end{itemize}

\subsection{On-axis location}
\label{sec:on-axis}

Let us first assume the on-axis placement of the experiment. We analyze how the number of long-lived FIPs is affected by changing the distance to the decay volume $l_{\text{min}}$, and its length $l_{\text{fid}}$ (the case of short-lived FIPs with $c\tau_{\text{FIP}}\gamma_{\text{FIP}}\lesssim l_{\text{min}}$ is discussed in Sec.~\ref{sec:upper-bound}).

\begin{figure*}[t]
    \centering
    \includegraphics[width=0.45\textwidth]{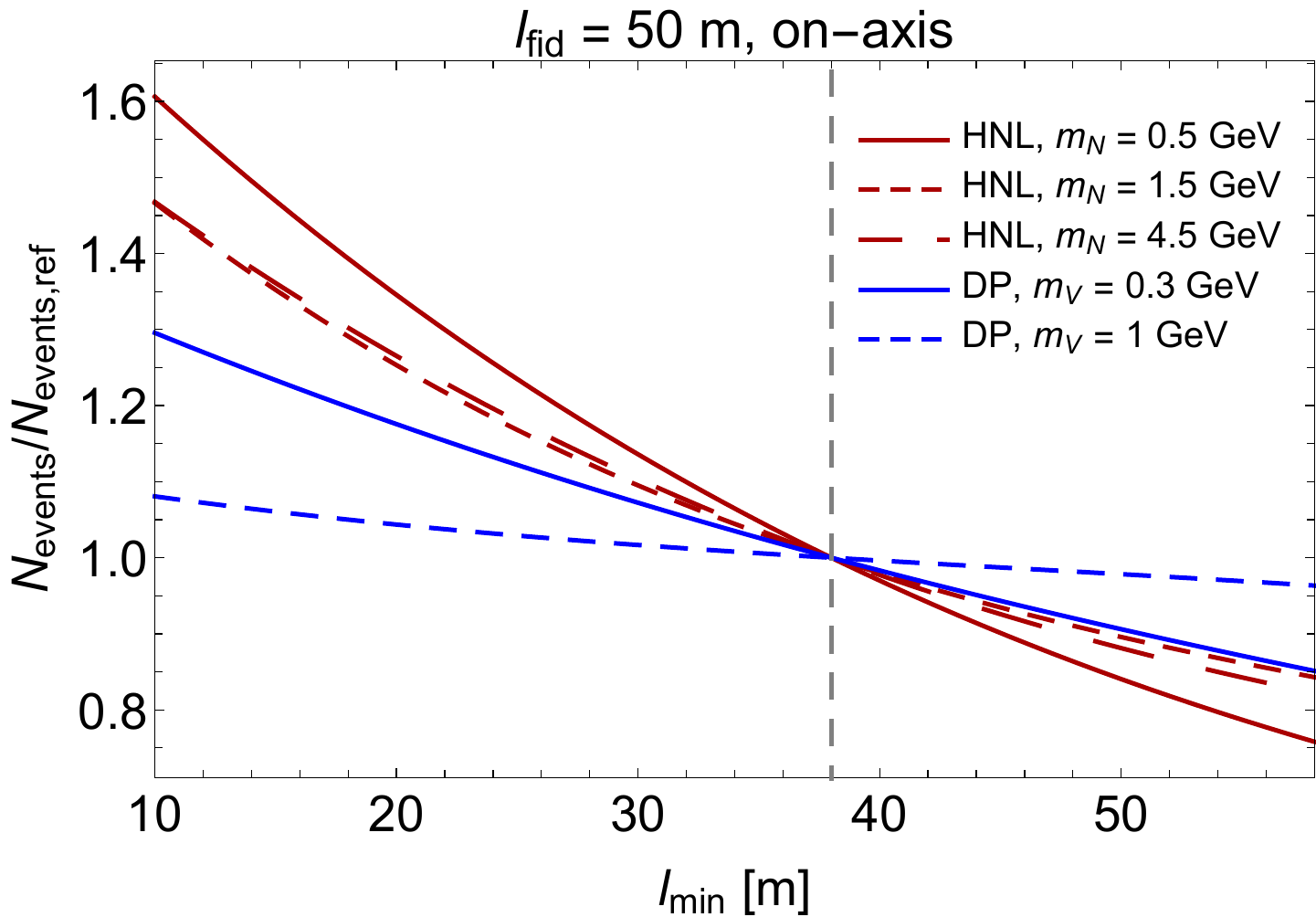}~\includegraphics[width=0.45\textwidth]{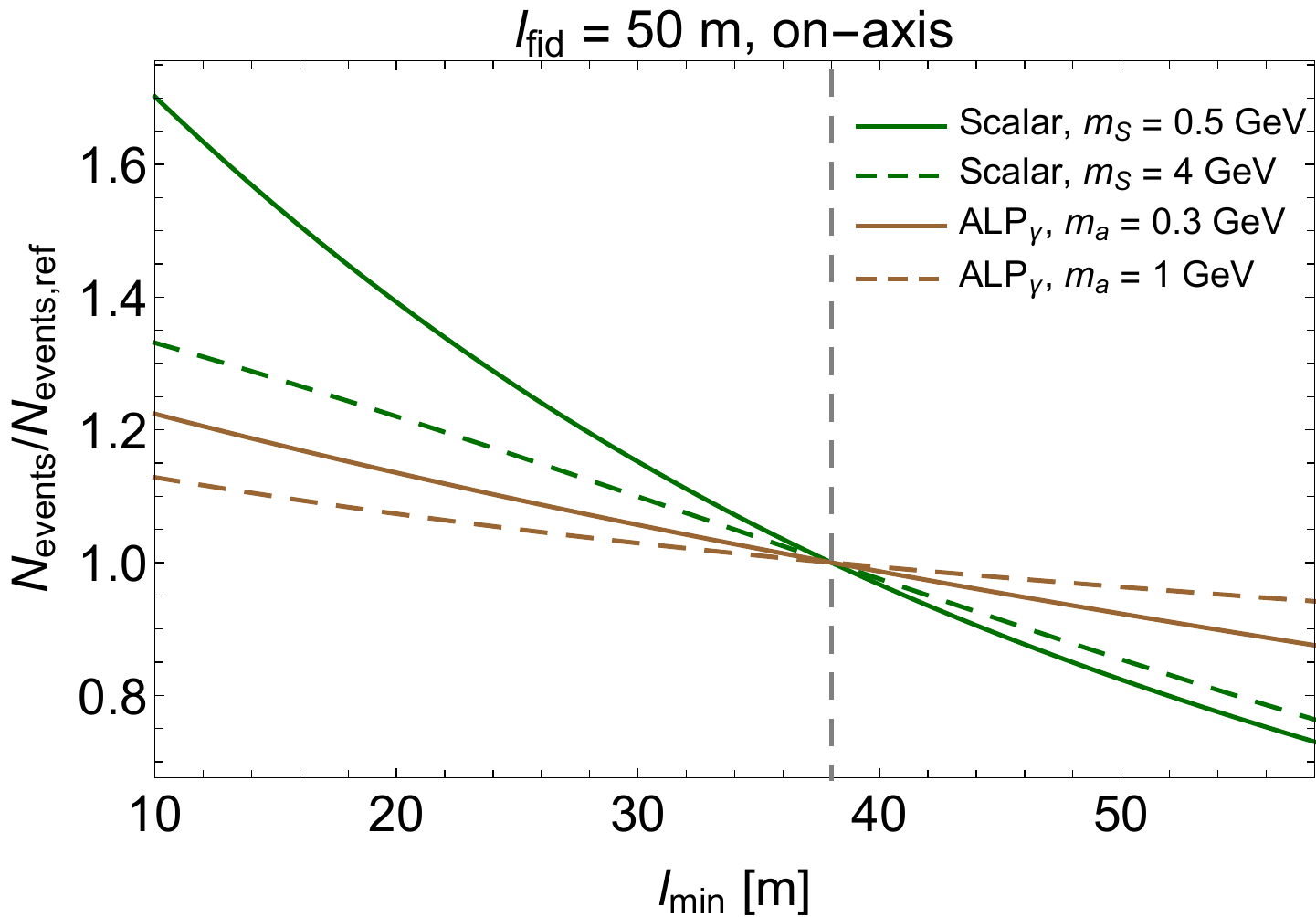} \\ \includegraphics[width=0.45\textwidth]{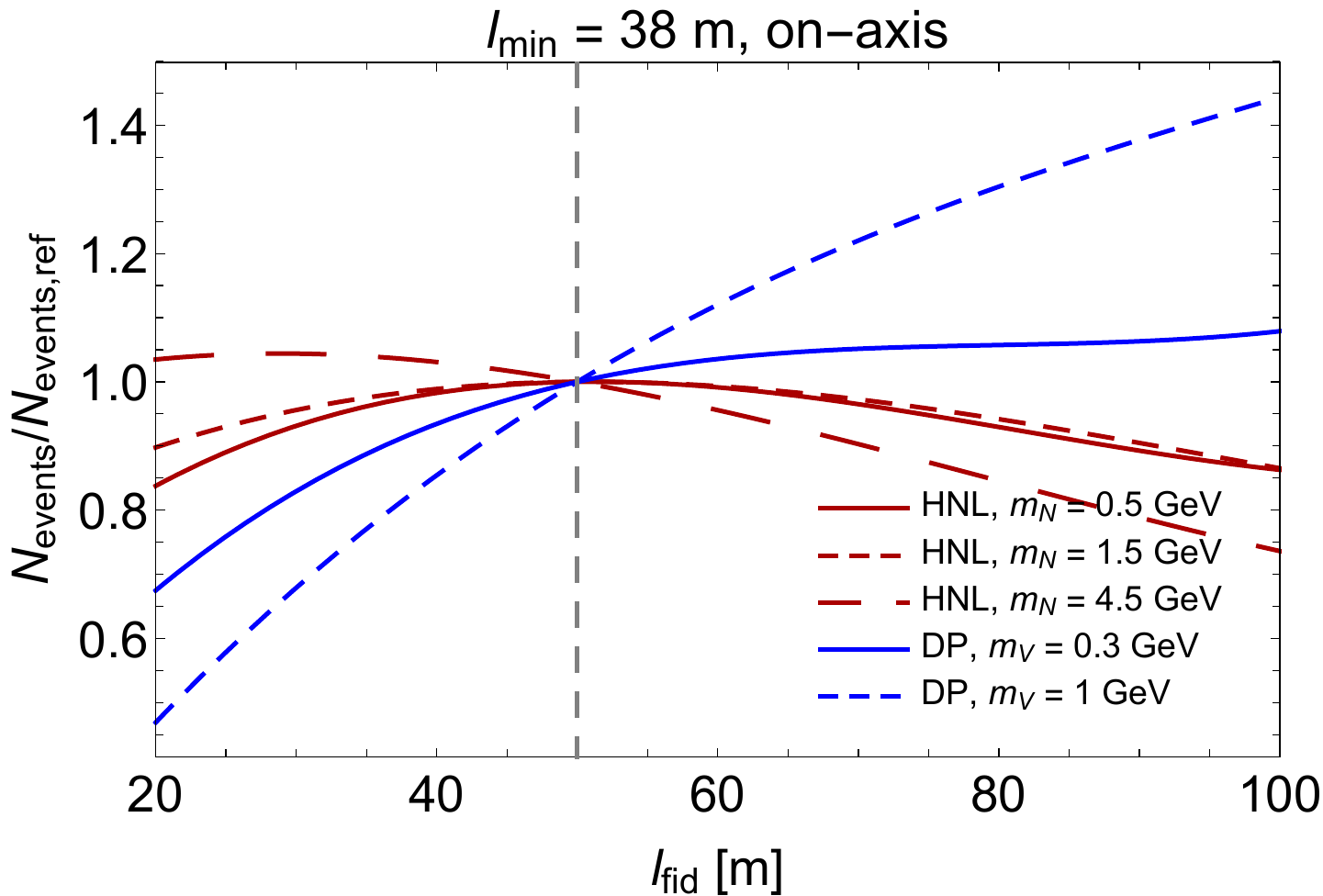}~\includegraphics[width=0.45\textwidth]{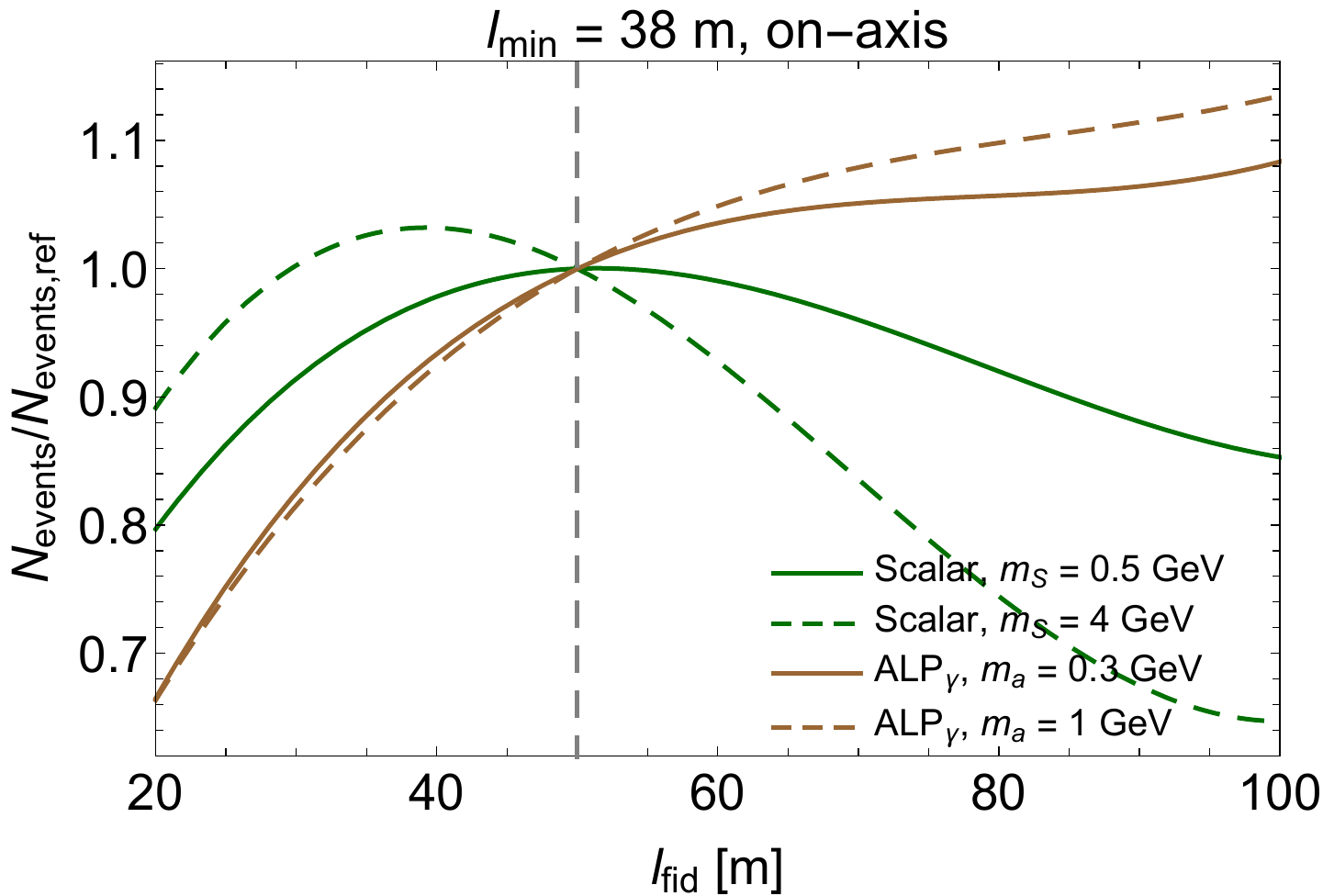}
    \caption{The behavior of the number of signal events of a beam dump on-axis experiment at the SPS at the lower bound of the sensitivity ($c\tau_{\text{FIP}}\langle\gamma_{\text{FIP}}\rangle\gg 100\text{ m}$, see Sec.~\ref{sec:qualitative-analysis} for details) under change of the distance to the decay volume $l_{\text{min}}$ (\textbf{top panels}) and its length $l_{\text{fid}}$ (\textbf{bottom panels}) for different models of FIPs. On the one hand, changing these parameters may have a significant impact for the backgrounds to be removed, the complexity of the setup, and costs. On the other hand, the maximal impact of these parameters on the number of events is small, $< \mathcal{O}(2)$, see text for details. Therefore, we conclude that the optimization of these parameters should be a subject of background considerations and costs rather than the maximization of the number of FIP events. The other parameters defining the experimental setup -- the transverse size of the decay volume and the detector dimensions -- are summarized in Table~\ref{tab:hypothetical-experiment-parameters}. For convenience, we normalize the number of events to the one for the configuration from Table~\ref{tab:hypothetical-experiment-parameters}.}
    \label{fig:on-axis}
\end{figure*}

The dependence of the number of events on $l_{\text{min}}$ is shown in Fig.~\ref{fig:on-axis} (top panels). We normalize all the values to the corresponding values of the SHiP-like experiment from Table~\ref{tab:hypothetical-experiment-parameters}. The main impact of $l_{\text{min}}$ is in defining the solid angle covered by the detector as seen from the target, 
\begin{equation}
\Omega_{\text{det-target}} = S_{\text{det}}/(l_{\text{min}}+l_{\text{fid}}+l_{\text{det}})^{2},
\label{eq:angle-target-det}
\end{equation}
and hence the fraction of FIPs $\epsilon_{\text{FIP}}$ pointing to the detector. If the FIP's angular distribution $df_{\text{FIP}}/d\Omega$  (Fig.~\ref{fig:FIPs-distributions}) is flat within the angles covered by the detector (the case of light HNLs, dark scalars, and ALPs), then $\epsilon_{\text{FIP}} \propto \Omega_{\text{det-target}}$. From Eq.~\eqref{eq:angle-target-det}, it follows that to increase $N_{\text{events}}$ for the SHiP-like configuration by a factor of two, it is necessary to decrease $l_{\text{min}}$ from $l_{\text{min}} = 38\text{ m}$ to $l_{\text{min}} \approx 8\text{ m}$. At the same time, if $df_{\text{FIP}}/d\Omega$ is collimated and falls at the boundaries (dark photon, ALPs, and heavy HNLs/dark scalars), the effect on $\epsilon_{\text{FIP}}$ is even smaller. On the other hand, the price of such a close placement would be a significant increase in the SM background.

The effect of $l_{\text{fid}}$ is less trivial, see the bottom panels of Figure~\ref{fig:on-axis}. It affects the product $\epsilon_{\text{FIP}}\times l_{\text{fid}}\times \epsilon_{\text{dec}}$. The impact on $\epsilon_{\text{FIP}}$ is similar to the $l_{\text{min}}$ case (Eq.~\eqref{eq:angle-target-det}). The second factor comes from the decay probability of long-lived FIPs. In addition, by increasing $l_{\text{fid}}$ and maintaining the aperture of the detector constant, $\epsilon_{\text{dec}}$ decreases, and vice versa. Indeed, $l_{\text{fid}}$ enters the solid angle covered by the detector as seen from the beginning of the decay volume, 
\begin{equation}
\Omega_{\text{det-fid}} = S_{\text{det}}/(l_{\text{fid}}+l_{\text{det}})^{2}
\end{equation}
If $l_{\text{fid}}$ is too large, the opening angle between the FIP's decay products $\Delta\theta_{\text{dec}}$ (Eq.~\eqref{eq:decay-angle}) becomes comparable with the detector size, and such events do not contribute. Hence, $l_{\text{fid}}\times \epsilon_{\text{dec}}$ remains constant under further increase of $l_{\text{fid}}$.

As a result, the product $\epsilon_{\text{FIP}}\times \epsilon_{\text{dec}}\times l_{\text{fid}}$ first scales linearly with $l_{\text{fid}}$, then reaches maximum where $\Delta\theta_{\text{dec}}$ becomes comparable with the detector size, and then decreases as a result of decreasing $\Omega_{\text{det-target}}$.

For the FIPs produced by decays of heavy mesons, the maximum is around $l_{\text{fid}} = 50\text{ m}$. The situation is different for the FIPs produced by decays of $\pi,\eta$, bremsstrahlung, or Primakov process. In these cases, either the FIPs have small masses or very large energies (Fig.~\ref{fig:FIPs-distributions}), which leads to a very large $\gamma_{\text{FIP}}$.  Hence, the suppression by $\epsilon_{\text{dec}}$ is not so severe, meaning that the number of events increases with $l_{\text{fid}}$ over a very large range.

\subsection{Off-axis location}
\label{sec:off-axis}

Let us now analyze the impact on the number of signal events when displacing the detector off-axis. As in the previous subsection, we first consider the configuration with the same dimensions and distance from the beam dump as in Table~\ref{tab:hypothetical-experiment-parameters}. We will consider only the parallel orientation of the decay volume and detector relative to the beamline, motivated by the limitations typically imposed by the infrastructure. Let us start with increasing the transverse displacement of the centre of the detector relative to the beamline, $r_{\text{displ}}$. We vary $r_{\text{displ}}$ from 0 to $5$ m. Note that $r_{\text{displ}}$ is not the same as the off-axis displacement of the side of the decay volume. In particular, for the considered configuration, the latter is $>0$ only if $r_{\text{displ}} >2\text{ m}$. For $r_{\text{displ}} =3\text{ m}$, the gap between the side of the decay volume and the beam axis is 1 m. 

\begin{figure*}[t]
    \centering
    \includegraphics[width=0.45\textwidth]{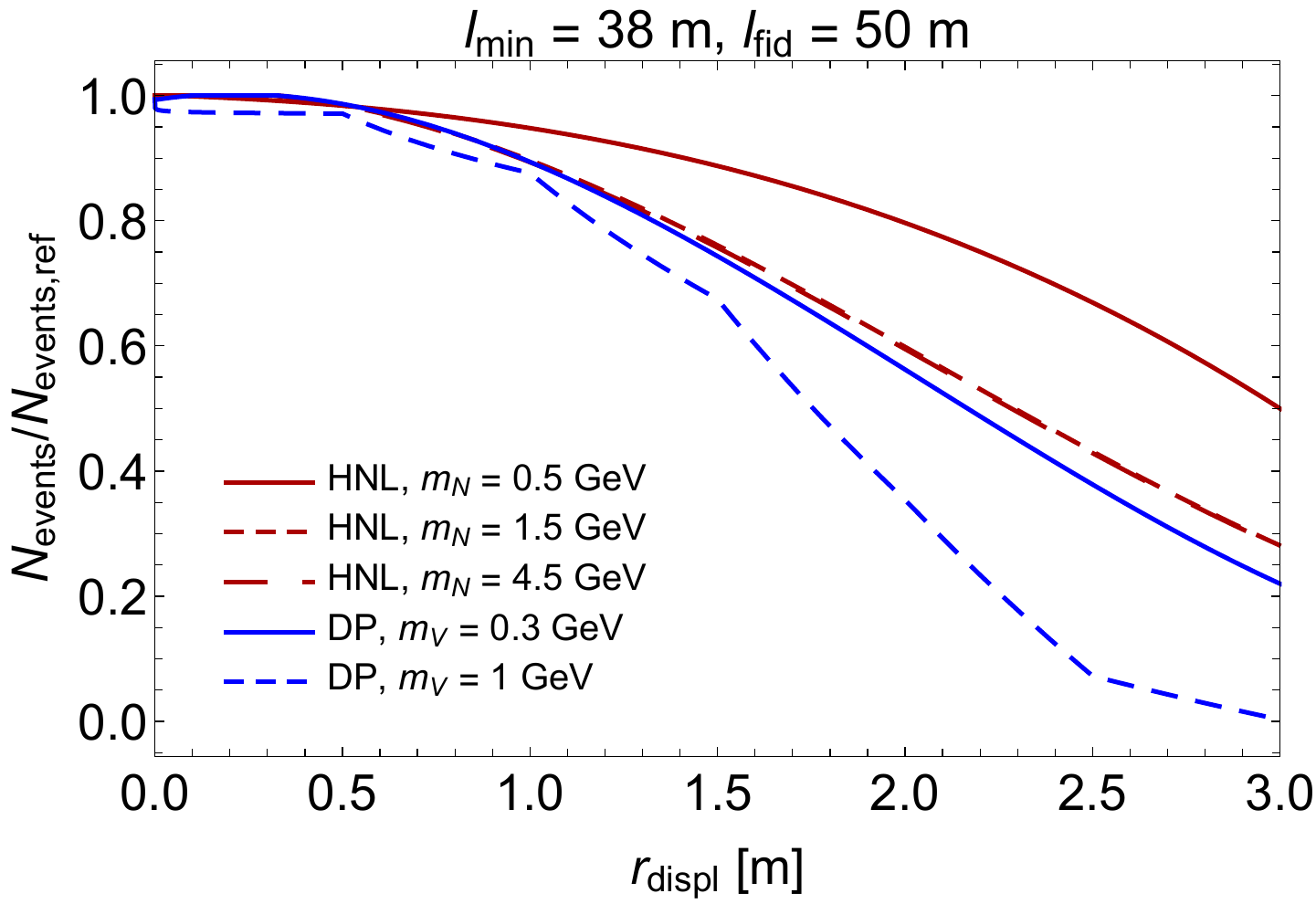}~\includegraphics[width=0.45\textwidth]{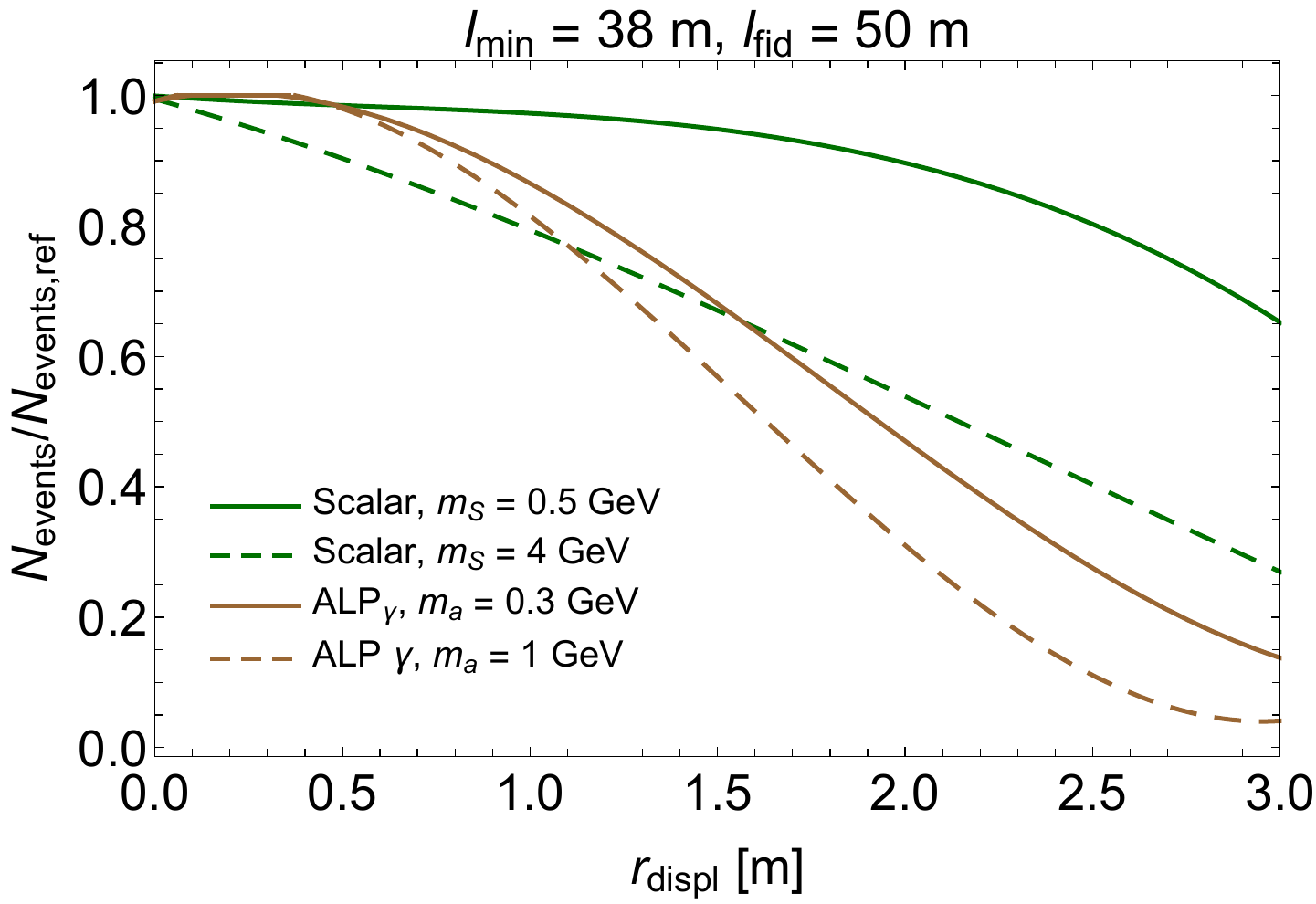}\\ \includegraphics[width=0.45\textwidth]{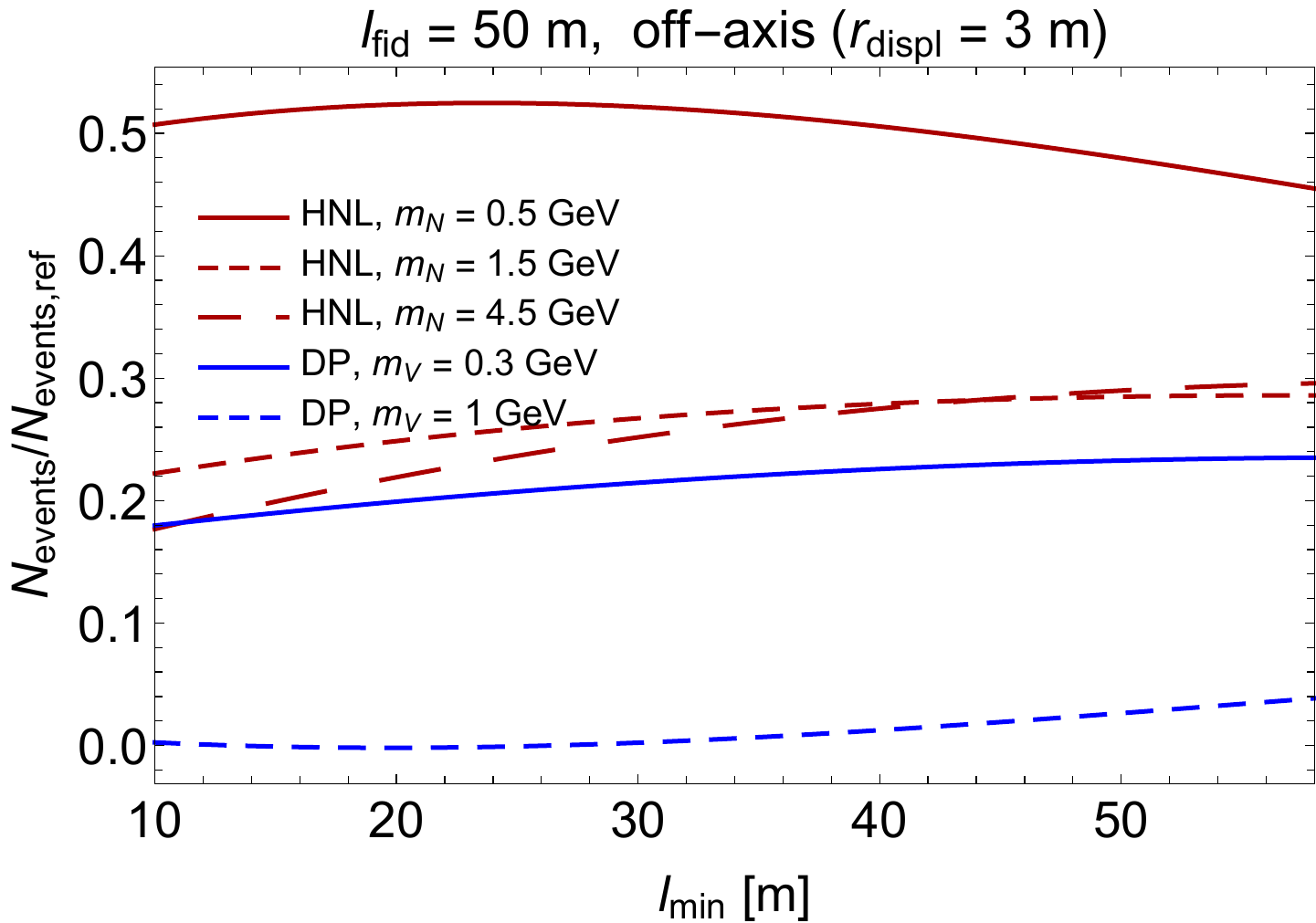}~\includegraphics[width=0.45\textwidth]{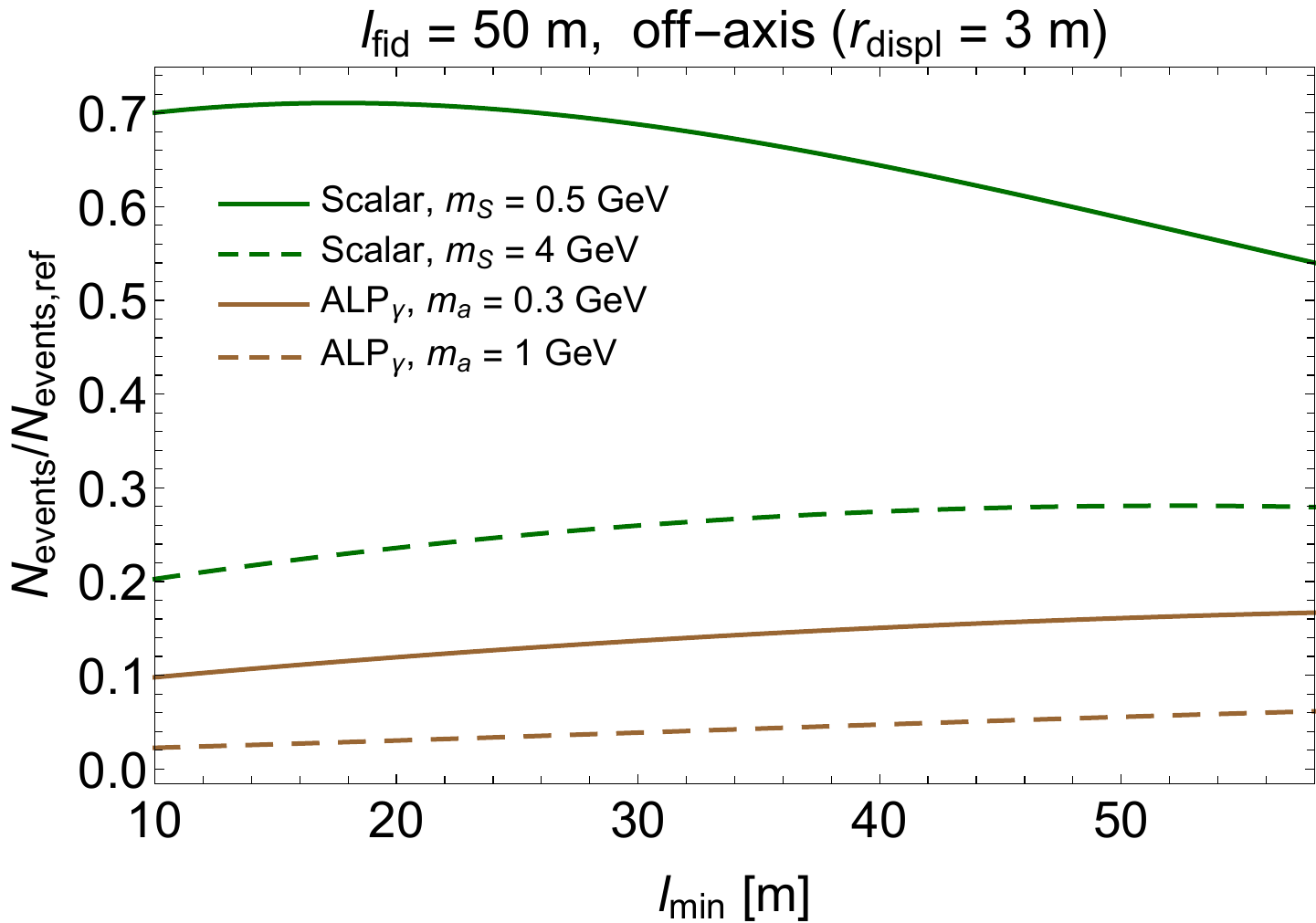}
    \caption{The behavior of the number of signal events of a beam dump experiment at the SPS at the lower bound of the sensitivity ($c\tau_{\text{FIP}}\langle\gamma_{\text{FIP}}\rangle\gg 100\text{ m}$, see Sec.~\ref{sec:qualitative-analysis}) assuming an off-axis placement of the centre of its decay volume parametrized in terms of the displacement $r_{\text{displ}}$. From the figures, we see that independently on the FIP type, by increasing $r_{\text{displ}}$, the number of events decreases. This results from the very forward-pointing FIP angular distribution that falls at large polar angles (Fig.~\ref{fig:FIPs-distributions}). Depending on the FIP, the decrease may be an order of magnitude or larger (\textbf{top panel}). It is impossible to compensate for this decrease by placing the experiment closer to the target (\textbf{bottom panel}): despite the increase of the solid angle covered by the detector, the minimal covered polar angle increases, which again results in a decrease of the FIP flux. The other parameters defining the experiment -- transverse dimensions of the decay volume, and detector dimensions -- are fixed as specified in Table~\ref{tab:hypothetical-experiment-parameters}. Note that $r_{\text{displ}}$ does not equal the off-axis displacement of the side of the decay volume. In particular, for the configuration considered, this displacement becomes non-zero only if $r_{\text{displ}} >2\text{ m}$. The displacement $r_{\text{displ}} = 3\text{ m}$ corresponds to 1 m gap between the side of the decay volume and beam axis. For convenience, we normalize the number of events to the one for the configuration specified in Table~\ref{tab:hypothetical-experiment-parameters}.}
    \label{fig:off-axis}
\end{figure*}

The dependence of the number of events is shown in Fig.~\ref{fig:off-axis}, top panels. For all FIP models considered, the number of events decreases with $r_{\text{displ}}$. The reason for this is that the FIP angular distribution decreases at the larger polar angles (Fig.~\ref{fig:FIPs-distributions}) covered with the off-axis displacement $r_{\text{displ}}$. In the case that the whole detector is placed off-axis ($r_{\text{displ}}>2\text{ m}$, such that the side of the decay volume is entirely away from the beam axis), we would not only shift the detector in the domain of large $\theta$, but also decrease the azimuthal coverage in the domain of small polar angles, which further reduces the FIP acceptance. Finally, the acceptance gets further suppressed by the shortening of the effective decay volume length for FIPs that enter from the side (Eq.~\eqref{eq:acceptance-qualitative}). 

For heavy FIPs, the drop is more significant, which is explained by more forward-pointing angular distributions and smaller decay acceptance. The decrease in $\epsilon_{\text{dec}}$ is due to the softer energy spectrum at large polar angles. It leads to an increase of the typical opening angle between the decay products $\theta \simeq 1/\gamma_{\text{FIP}}$ (Eq.~\eqref{eq:decay-angle}), and hence a decrease in $\epsilon_{\text{dec}}$.

It is also interesting to compare the effect of shortening $l_{\text{min}}$ in the case of off-axis and on-axis placements (Fig.~\ref{fig:on-axis} and discussion therein). The behavior of the number of events is illustrated in Fig.~\ref{fig:off-axis}, bottom panels. Unlike the on-axis case, the number of events tends to grow if {\it increasing} $l_{\text{min}}$, which is again a result of the very forward-pointing angular distributions and decreasing $\epsilon_{\text{dec}}$ at the off-axis locations.

\subsection{Upper bound of the sensitivity for on-axis and off-axis}
\label{sec:upper-bound}
Finally, let us also examine the role of the geometric parameters in determining the potential of the experiment to search for FIPs with large couplings, for which the typical FIP decay length is smaller than the distance to the decay volume, $c\tau_{\text{FIP}}\langle\gamma_{\text{FIP}}\rangle \lesssim l_{\text{min}}$.

At the upper bound, the number of events is proportional to the following integral (see Appendix~\ref{app:numerics}):
\begin{multline}
    N_{\text{events}}^{\text{upper bound}} \propto \epsilon = \int dL \int dE_{\text{FIP}} \\ \frac{1}{c\tau_{\text{FIP}}\gamma_{\text{FIP}}}\exp\left[-\frac{L}{c\tau_{\text{FIP}}\gamma_{\text{FIP}}}\right]f^{l_{\text{min}}}_{E_{\text{FIP}},L},
    \label{eq:upper-bound} 
\end{multline}
where $L$ is the modulus of the FIP decay position, and $f^{l_{\text{min}}}_{E_{\text{FIP}},L} \equiv \bigg\langle \frac{df_{\text{FIP}}}{dE_{\text{FIP}}}\epsilon_{\text{dec}}\bigg\rangle_{\theta}$ is the FIP distribution averaged over the angular coverage of the detector. The integral~\eqref{eq:upper-bound} effectively plays the role of geometric acceptance in the case of short-lived FIPs. It is sensitive to the high-energy tail of the FIP distribution ($E_{\text{FIP}}>\langle E_{\text{FIP}}\rangle$).

We will start with the simpler case of the on-axis configuration. In this case,  $f^{l_{\text{min}}}_{E_{\text{FIP}},L}$ depends weakly on $l_{\text{min}}$ and on $L$: $f^{l_{\text{min}}}_{E_{\text{FIP}},L} \approx f_{E_{\text{FIP}}}$. Indeed, independently of $l_{\text{min}}$, the detector covers the far-forward domain, which determines the high-energy tail of the distribution function. As a result, the only impact of decreasing $l_{\text{min}}$ comes from decreasing $L_{\text{min}}\approx l_{\text{min}}$. Namely, the whole integral is saturated around its value $L_{\text{min}}$:
\begin{equation}
    \epsilon \approx \int dE_{\text{FIP}}\exp\left[-\frac{l_{\text{min}}}{c\tau_{\text{FIP}}\gamma_{\text{FIP}}}\right]f_{E_{\text{FIP}}}
\end{equation}
Up to logarithmic corrections, at a fixed mass, the upper bound of the sensitivity, i.e., the smallest lifetimes that may be probed, scales as $\tau_{\text{FIP}}^{\text{upper}}\propto l_{\text{min}}^{-1}$. 

Let us now consider the off-axis case, concentrating on the case where the whole detector lies off-axis. For the detector dimensions in Table~\ref{tab:hypothetical-experiment-parameters}, this case would correspond to the transverse displacement of the centre of the detector $r_{\text{displ}}>3\text{ m}$. In this case, the situation is less trivial. First, the FIP energy spectrum becomes softer (remind Fig.~\ref{fig:FIPs-distributions}) at large polar angles. As a result, the value of the $E_{\text{FIP}}$ integral in Eq.~\eqref{eq:upper-bound} at fixed $L$ decreases compared to the off-axis case. Second, the FIPs pointing to the closest (to the beamline) part of the detector enter the decay volume from the side (Fig.~\ref{fig:acceptance-illustration}). This also decreases $f^{l_{\text{min}}}_{E_{\text{FIP}},L}$. This behavior is in tension with the exponent in Eq.~\eqref{eq:upper-bound}: small values of $L$ close to $L_{\text{min}}$ correspond to maximal polar angle and hence smaller energies/acceptance. This destructive interplay destroys the positive impact of decreasing $l_{\text{min}}$ on the potential to probe short FIP lifetimes. As a result, the fully off-axis configurations have a lower number of signal events than the on-axis, even given much smaller $l_{\text{min}}$.

\begin{figure}[!h]
    \centering
    \includegraphics[width=0.45\textwidth]{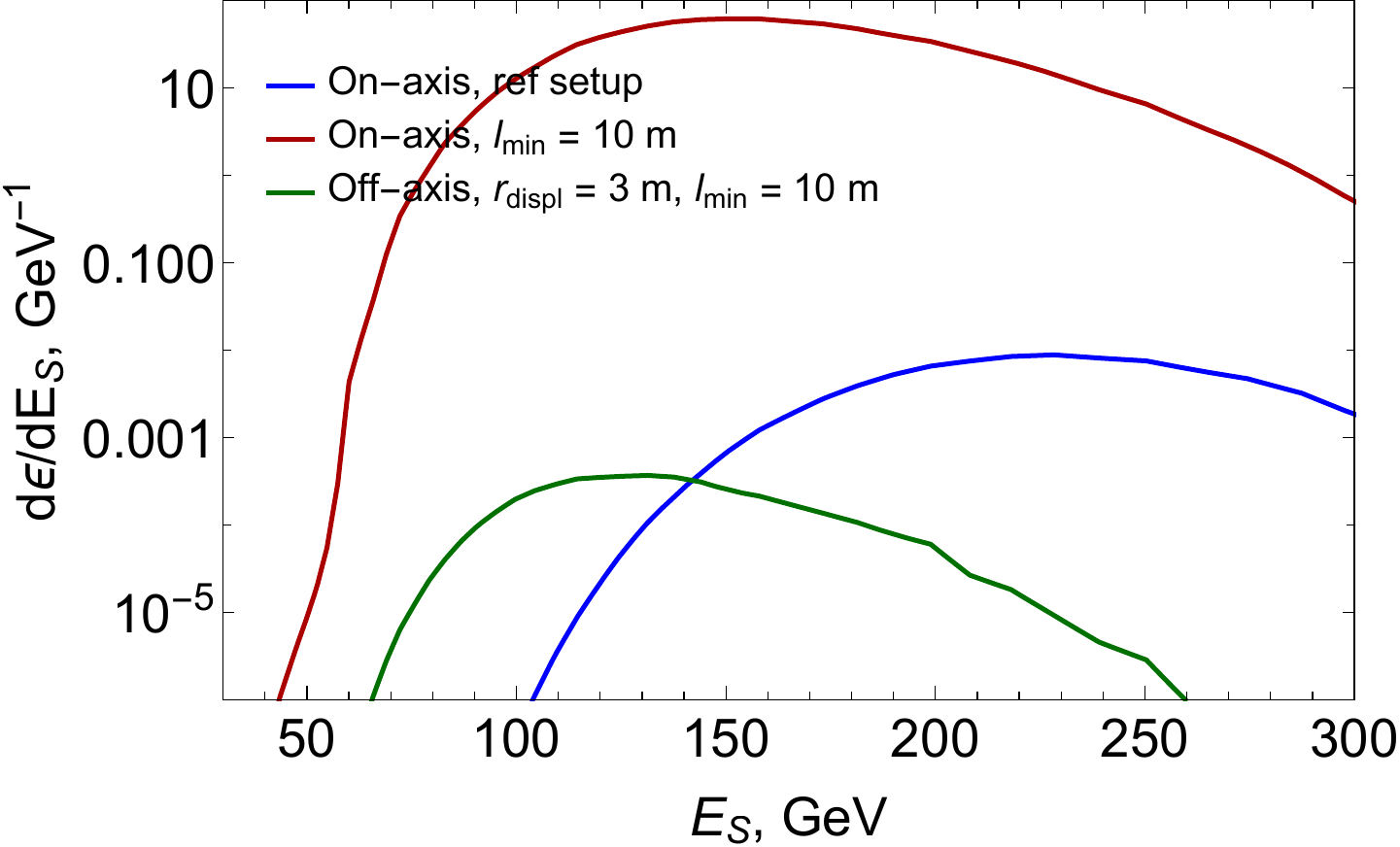}
    \caption{The energy distribution of the function~\eqref{eq:energy-spectrum-upper-bound} determining the number of events at the upper bound of the sensitivity ($c\tau_{\text{FIP}}\langle\gamma_{\text{FIP}}\rangle\lesssim l_{\text{min}}$, where $l_{\text{min}}$ is the distance from target to the decay volume). Its shape and normalization depend on the value of the distance from the target to the decay volume $l_{\text{min}}$ and the high-energy tail of FIPs within the acceptance for the given experiment. If decreasing $l_{\text{min}}$ for the on-axis placement of the experiment, the high-energy tail would remain unchanged; as a result, we would increase the event rate. This may not be the case for an off-axis placement: the energy spectrum may become much softer and compensate for the decrease of $l_{\text{min}}$ (see text for details). To illustrate these points, we consider a dark scalar with mass $m_{S} = 3\text{ GeV}$ and lifetime $c\tau_{S} = 0.05\text{ m}$ at three various experimental setups at SPS: the configuration from Table~\ref{tab:hypothetical-experiment-parameters}; the same configuration but with $l_{\text{min}} = 10\text{ m}$; and the off-axis experiment with $l_{\text{min}} = 10\text{ m}$, the displacement of the lower edge of its decay volume from the beamline of $1\text{ m}$, and the decay volume length $l_{\text{fid}} = 20\text{ m}$. The number of events at the closer on-axis experiment is larger, while at the off-axis experiment, it is smaller since scalars have smaller energies.}    
    \label{fig:upper-bound}
\end{figure}

To illustrate these qualitative arguments, let us consider three setups: the on-axis experiment from Table~\ref{tab:hypothetical-experiment-parameters}, the same on-axis experiment but with $l_{\text{min}} = 10\text{ m}$, and finally an off-axis experiment with $l_{\text{min}} = 10\text{ m}$ and $r_{\text{displ}} = 3\text{ m}$. For the FIPs, we will consider dark scalars with mass $m_{S} = 3\text{ GeV}$ and lifetime $c\tau_{S} = 5\text{ cm}$, which corresponds to the upper bound of the sensitivity.

The behavior of the integrand of~\eqref{eq:upper-bound},
\begin{equation}
\frac{d\epsilon}{dE_{\text{FIP}}} = \int dL \frac{1}{c\tau_{\text{FIP}}\gamma_{\text{FIP}}}\exp\left[-\frac{L}{c\tau_{\text{FIP}}\gamma_{\text{FIP}}}\right]f^{l_{\text{min}}}_{E_{\text{FIP}},L},
\label{eq:energy-spectrum-upper-bound}
\end{equation}
for the three setups is shown in Fig.~\ref{fig:upper-bound}. Obviously, the on-axis setup with $l_{\text{min}} = 10\text{ m}$ has the largest flux of FIPs. However, despite the fact that the off-axis decay volume is located $\sim 4$ times closer to the beam dump than the on-axis setup with $l_{\text{min}} = 38\text{ m}$, the off-axis setup has a $\simeq 30$ times lower value of the total integral~\eqref{eq:upper-bound}, following from the factors described above.

\section{Comparison between the experiment proposals under study at the CERN SPS}
\label{sec:comparison-proposals}

In this section, we make a comparison of the physics yields between the three experiment proposals that are currently being considered for implementation in the ECN3 beam facility at CERN's SPS accelerator, HIKE~\cite{CortinaGil:2839661}, SHADOWS~\cite{Alviggi:2839484}, SHiP~\cite{Aberle:2839677}. All proposals are based on a similar detector setup in that they conceptually consist of large decay volumes followed by spectrometers and particle identification, together with various veto systems. HIKE is primarily a kaon experiment located on-axis. It requires a specialised beam setup with a kaon target, a secondary beam line for the kaon selection, and an absorber of copper/iron for the remaining proton beam and secondary hadrons from the kaon target. HIKE's distance to the kaon target is defined by the optimisation for the kaon physics programme, resulting in a relatively large distance between HIKE's decay volume and the absorber. The kaon physics optimisation imposes limitations on the maximum beam intensity that is due to both the secondary beam line setup and the detector, effectively four times lower than it would be for SHiP. HIKE is also proposed to partially operate in beam-dump mode for FIP physics. In this mode, the kaon target is moved aside to let the proton beam of the same intensity be directly dumped on the absorber. With the very small solid angle coverage, HIKE has mainly sensitivity to dark photons and ALPs with photon coupling. SHADOWS is an off-axis experiment that would be located alongside HIKE's kaon beam, downstream of the absorber, with the detector covering angles $\theta \gtrsim 30\text{ mrad}$ (Fig.~\ref{fig:FIPs-distributions}). SHADOWS' distance to the proton absorber is defined by the infrastructure around the absorber, shielding requirements, a muon sweeper, and the subsequent beam line elements. SHADOWS would operate together with HIKE in beam-dump mode, with SHADOWS searching for the FIPs flying off-axis and HIKE for those produced in the far-forward region. In this respect, it is expected that the beam time for HIKE and SHADOWS is split between periods of kaon physics and beam-dump physics. SHiP is instead a dedicated on-axis experiment with the detector located as close as possible to a compact target station housing a target of molybdenum/tungsten that is optimised for FIP physics. SHIP's distance to the target is defined by a hadron stopper with minimum depth, and a specialised magnetic muon deflector that sweeps the muon flux away from the fiducial volume. SHIP's location allows it to cover all the FIP production modes at the SPS.

\begin{figure*}[t]
    \centering
  \includegraphics[width=0.45\textwidth]{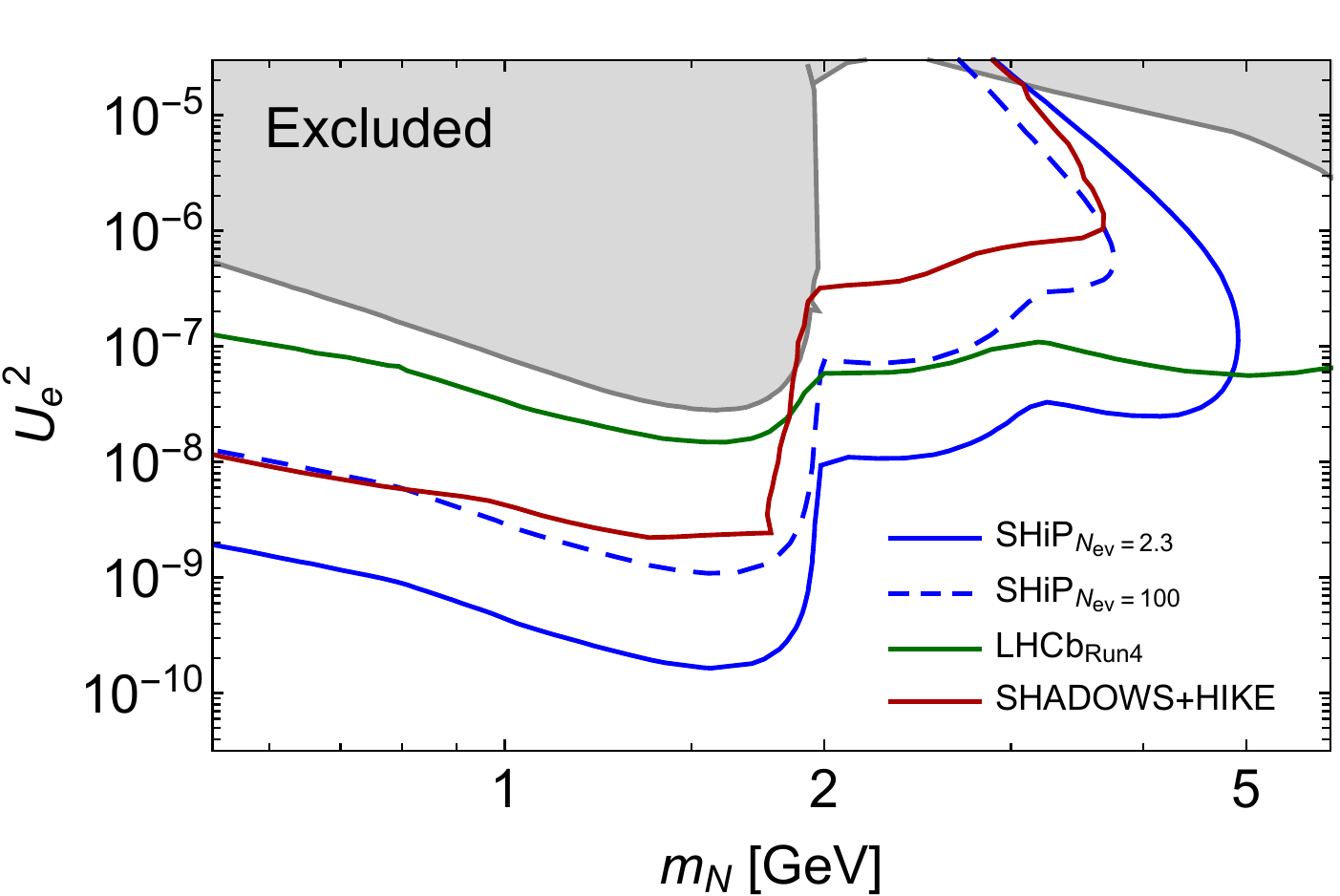}~\includegraphics[width=0.45\textwidth]{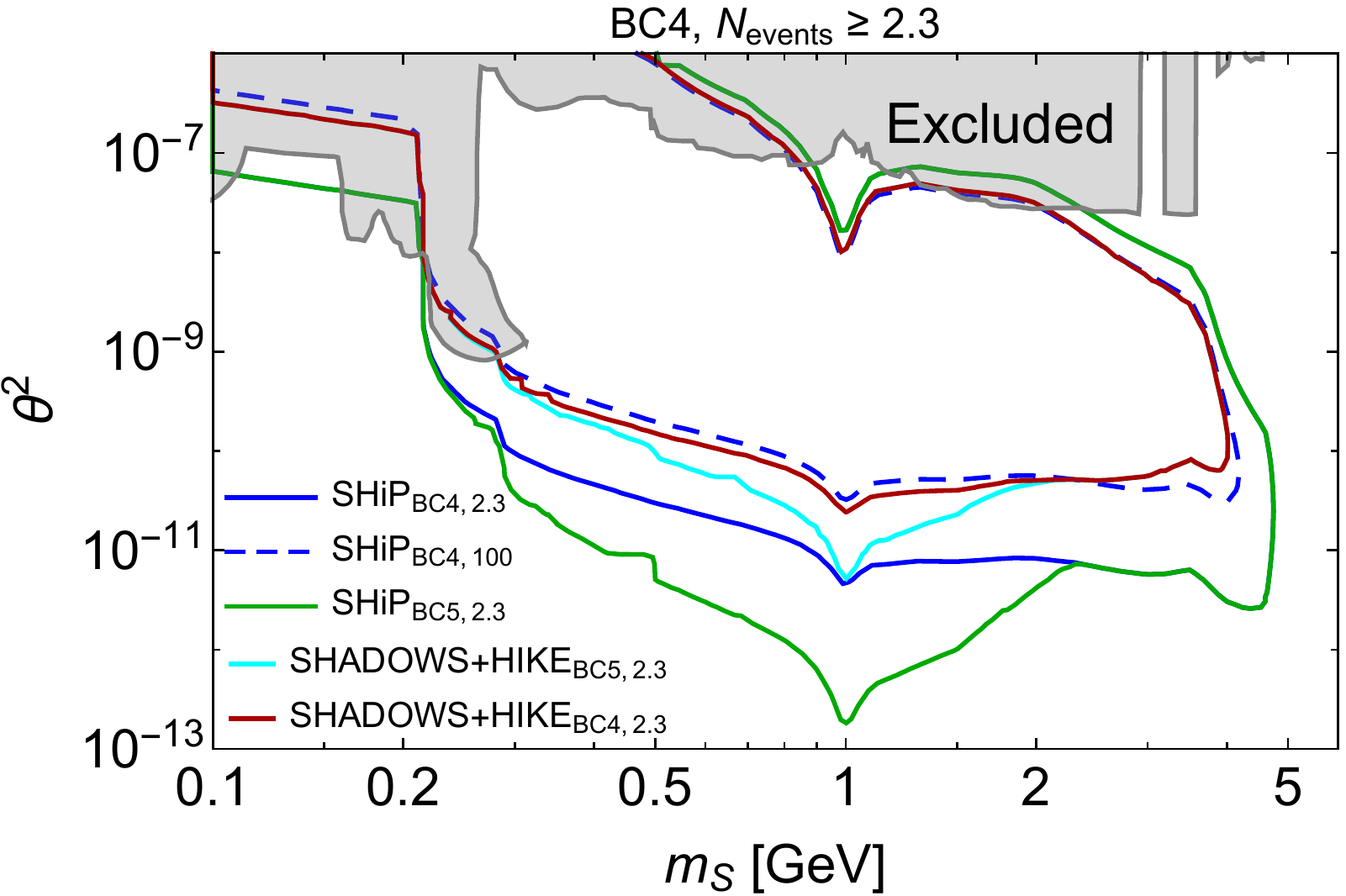}\\\ \includegraphics[width=0.45\textwidth]{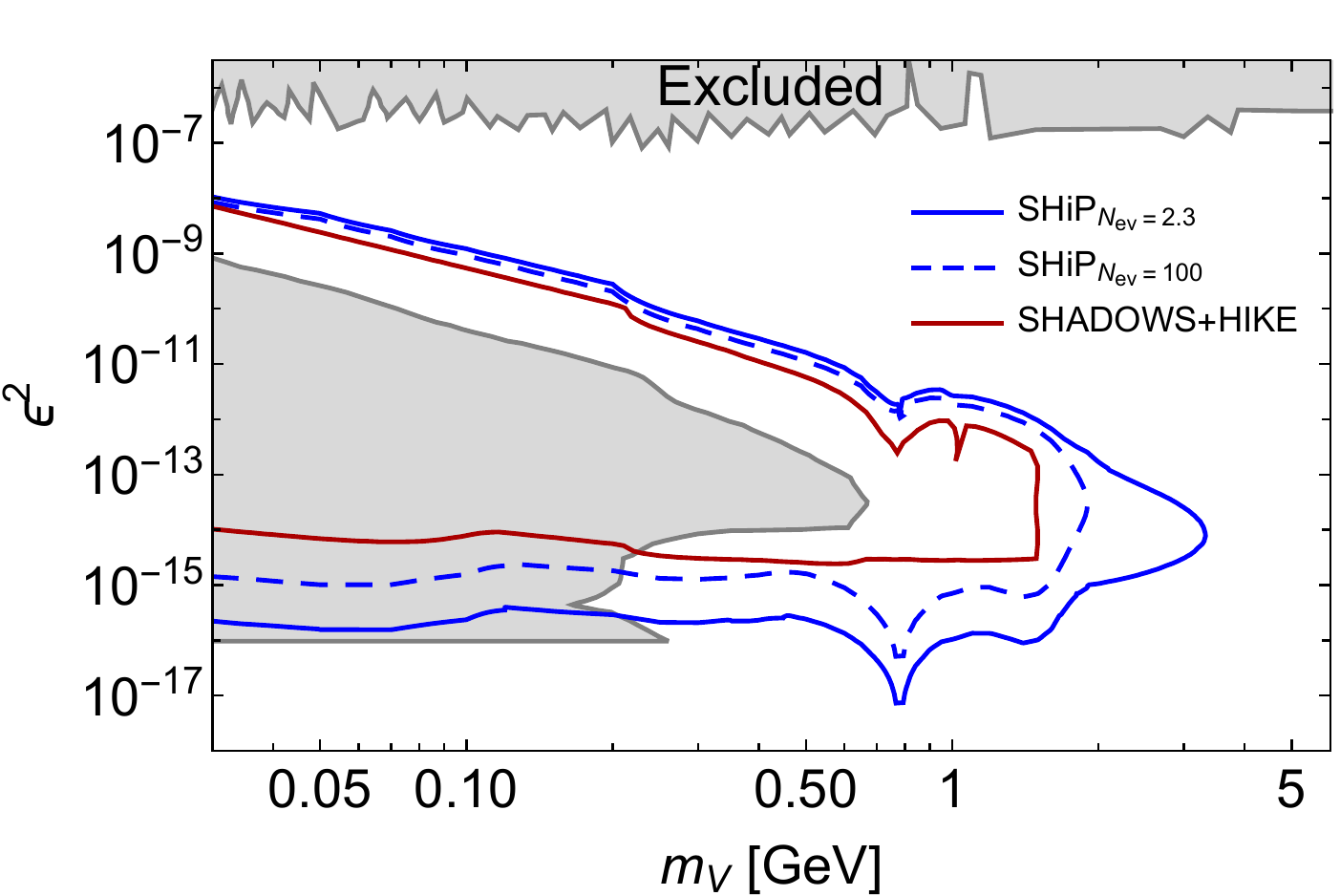}~\includegraphics[width=0.45\textwidth]{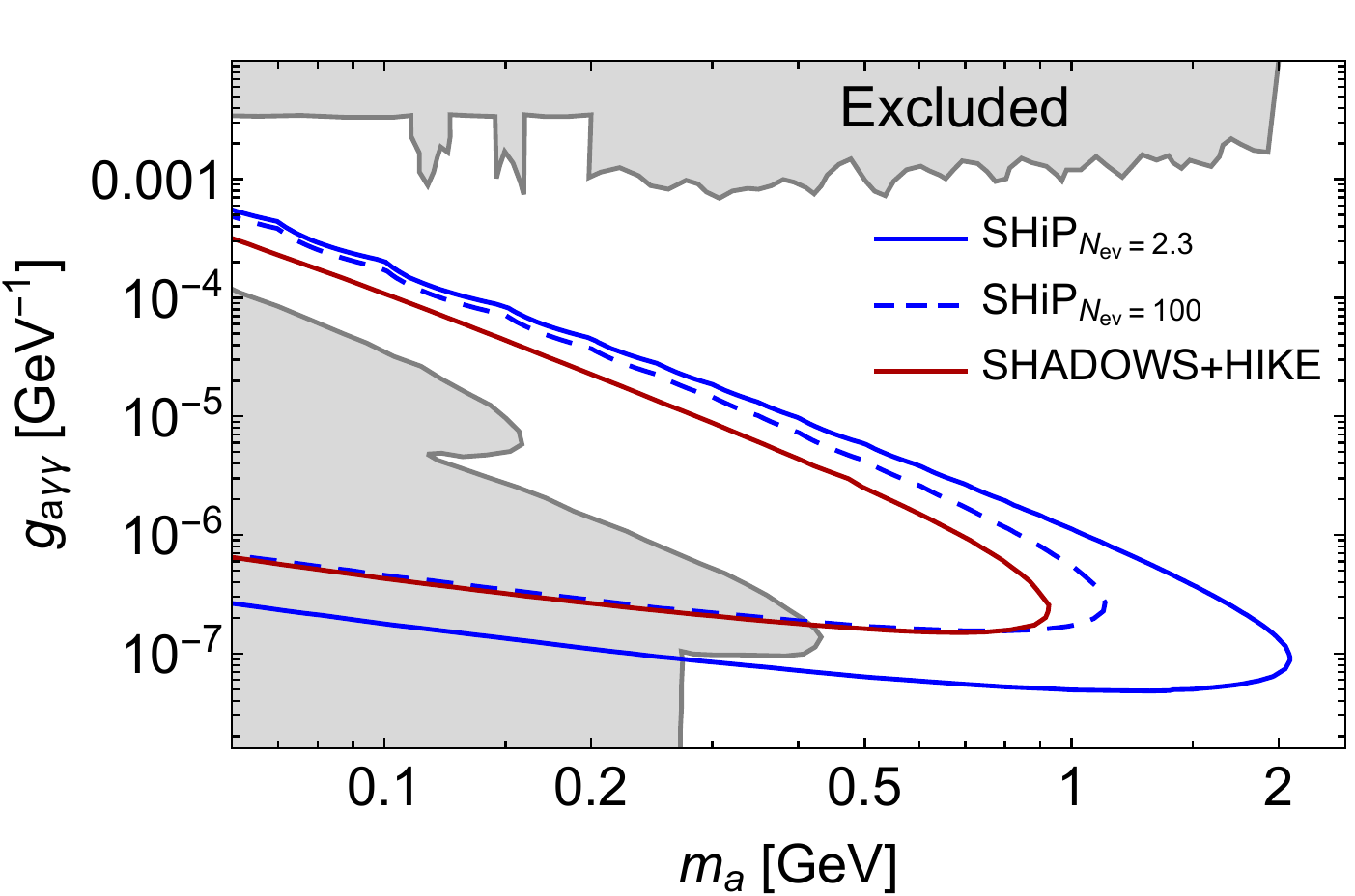}
    \caption{Comparison of the potential of the HIKE, SHADOWS, and SHiP proposals to explore the parameter space of FIPs in beam-dump mode. The FIP physics is here illustrated with HNLs (\textbf{top left panel}), dark scalars with the mixing coupling and with the quartic coupling fixed by $\text{Br}(h\to SS) = 10^{-3}$ (\textbf{top right panel}), dark photons (\textbf{bottom left}), and ALPs with the coupling to photons (\textbf{bottom right}). For all the models except for dark scalars, we use the 90\% CL combined sensitivity of SHADOWS and HIKE$_{\text{dump}}$ from the LoIs~\cite{Alviggi:2839484,CortinaGil:2839661}, while for dark scalars we report our own estimates based on the SHADOWS configuration specified in the LoI (see Appendix~\ref{app:numerics} for detail). For SHiP, we show two curves: the 90\% CL sensitivity, and the domain $N_{\text{events}} >100$, where it may be possible to determine properties of FIPs such as their mass and decay branching ratios. The combined impact of the lower number of protons on the target and the non-optimal placement for SHADOWS and HIKE lead to a significant limitation on their physics potential. Their exclusion domain lies within the FIP identification domain of SHiP. Moreover, in the case of FIPs produced by decays of $B$ mesons (such as e.g. HNLs, dark scalars, and ALPs coupled to fermions), the reach of SHADOWS and HIKE may be overcome by future searches at LHCb~\cite{LHCP:2022} with triggers allowing the use of the muon stations as trackers. We show the sensitivity of such types of searches in the case of HNLs.}
    \label{fig:summary}
\end{figure*}

In Fig.~\ref{fig:summary} we show the 90\% CL sensitivities of HIKE$_{\text{dump}}$ + SHADOWS and of SHIP to HNLs, dark scalars, dark photons, and ALPs with the coupling to photons. In addition, for SHiP, we include the iso-contours corresponding to $N_{\text{events}}=100$. Such a large number of events allows not only to establish the existence of a new particle but also identify its properties  such as branching ratios of various decay channels, precise mass, etc. Details of the sensitivity estimates are described in Appendix~\ref{app:numerics}. We see that the lower bound of the sensitivity of SHADOWS+HIKE$_{\text{dump}}$ is close to the 100 events line of SHiP. This may be easily understood if using Eq.~\eqref{eq:Nevents-approx}.
Namely, for SHADOWS, the product $N_{\text{PoT}}\times l_{\text{fid}}$ is 10 times smaller than at SHiP. The rest of the suppression in the number of events at SHADOWS comes from the off-axis placement parallel to the beamline, significantly decreasing $\epsilon_{\text{FIP}}\times \langle p^{-1}_{\text{FIP}}\rangle$ -- by a factor $1/3-1/20$ depending on the FIP mass. For HIKE, the product $N_{\text{PoT}}\times l_{\text{fid}}$ is only twice smaller than at SHiP. However, its detector covers $\sim 20$ times smaller solid angle as seen from the target and the beginning of the decay volume, which results in a much lower overall acceptance.

Moreover, despite the fact that SHADOWS is placed much closer to the target, which could naively mean that it should be able to probe larger couplings of FIPs, Fig.~\ref{fig:summary} shows that this is not the case. One of the reasons for this is that FIPs flying off-axis have a much softer energy spectrum and hence shorter typical decay length (see Fig.~\ref{fig:off-axis} and discussions therein). An additional reason is that the most energetic FIPs enter the decay volume from the side, resulting in a shorter flight path within the decay volume and thus suppressing the fraction of their decays inside the decay volume.

\section{Conclusions}
\label{sec:conclusions}

Future searches for new physics beyond the Standard Model are without a doubt in need of a diverse approach and experiments with complementary sensitivities to different classes of models, ranging both mass and coupling scales.  A theoretically and experimentally attractive case for new physics that is largely unexplored but within reach at current facilities, consists of particles with masses below the electroweak scale that may be produced at accelerators in decays of SM particles. Their couplings should be small in order to avoid existing experimental constraints. Hence, they are often called feebly-interacting particles, or FIPs.

FIPs are now being actively searched for at the LHC and will be searched for at future collider experiments. However, this type of search has limitations in probing the parameter space of long-lived FIPs in the mass range $m_{\text{FIP}}\lesssim 5-10\text{ GeV}$, see Fig.~\ref{fig:light-FIPs-colliders}, mainly because of strong backgrounds and too short decay volumes as compared to the typical decay length of light FIPs. Beam dump-like experiments, characterized by the possibility of extremely high luminosity at relatively high energies, and effective coverage of the production and decay acceptance, are the perfect setup to generically explore the "coupling frontier" and the case of FIPs with mass below the B meson mass. Beam dump experiments can be equipped with long decay volumes and be located at some distance from the target to accommodate absorbers and deflectors of SM particles, as well as veto systems, to reduce the backgrounds. The best placement for such a FIP facility is the SPS accelerator at CERN, where the existing infrastructure and the currently available proton yield of up to $4\times10^{19}$ protons per year at 400\,GeV make it possible to implement and operate a-state-of-the-art experiment at relatively low costs. There are currently three experiment proposals being considered for implementation in the ECN3 beam facility at CERN's SPS accelerator, HIKE~\cite{CortinaGil:2839661}, SHADOWS~\cite{Alviggi:2839484}, SHiP~\cite{Aberle:2839677}. Their respective objectives and layouts are briefly summarised in Sec.~\ref{sec:comparison-proposals}.

In order to determine the optimal experimental geometry, we have made an in-depth study of the dependence of the number of FIP signal events at the lower (Secs.~\ref{sec:on-axis},~\ref{sec:off-axis}) and the upper (Sec.~\ref{sec:upper-bound}) bounds of the sensitivity, as a function of the length of the decay volume, the distance from the target, and the transverse displacement with respect to the beamline, and other parameters. We have performed the analysis for several classes of models of FIPs (``portals'') with different production channels and decay modes, see Sec.~\ref{sec:fip-phenomenology}. Given that all the proposals claim to reach a background-free regime, we do not consider backgrounds specifically other than the constraints that such assumptions impose on the geometric parameters.

In particular, we analyzed the effect of displacing the detector off-axis for different portals. Generically, it leads to a decrease in the number of events for two specific reasons: first, the angular flux of FIPs decreases at large angles (Fig.~\ref{fig:FIPs-distributions}, and second, it causes geometric shortening of the effective length of the decay volume along trajectories pointing to the detector (Fig.~\ref{fig:acceptance-illustration}). The impact of the off-axis placement depends on the dominant FIP production channel, i.e., by the decays of mesons (such as for HNLs, ALPs, and dark scalars), or directly in proton-target collisions (dark photons). In the former case, shifting the decay volume entirely to the side of the beam axis such that it covers polar angles $\theta\gtrsim 10\text{ mrad}$, leads to a loss of up to a factor of five in the number of events at the lower bound, depending on the FIP mass, as compared to the reference configuration in Table~\ref{tab:hypothetical-experiment-parameters}. In the latter case, the same decrease in yield is seen as soon as the side of the decay volume wall gets off-axis. Practically, this means that the off-axis configurations have no sensitivity to dark photons.
  
In contrast, we have seen that changing the distance from the target to the decay volume, and its length, for the on-axis configuration does not affect the number of events at the lower bound by more than a factor of two (Fig.~\ref{fig:on-axis}) over a broad range of parameters. On the other hand, significantly decreasing the distance to the target affects the complexity of the experiment and background suppression, and hence the cost. Therefore, optimizing the on-axis configuration is rather a subject of minimizing background and cost.

Finally, we have applied the analysis to the three ECN3 proposals (Sec.~\ref{sec:comparison-proposals}). In Fig.~\ref{fig:summary}, we show the 90\% CL sensitivities of HIKE$_{\text{dump}}$ + SHADOWS and of SHIP to HNLs, dark scalars, dark photons, and ALPs with the photon coupling. We also include the projection of the sensitivity of the LHC experiments, taking LHCb as an example, for the case that new triggers will be developed, allowing, e.g., the use of the muon chambers as trackers for FIPs independently of their production channel~\cite{LHCP:2022}. In this case, the sensitivity of SHADOWS+HIKE may be limited in the domain of heavy FIPs with $m\gtrsim 2\text{ GeV}$.

\begin{figure*}[t]
    \centering
    ~\includegraphics[width=0.45\textwidth]{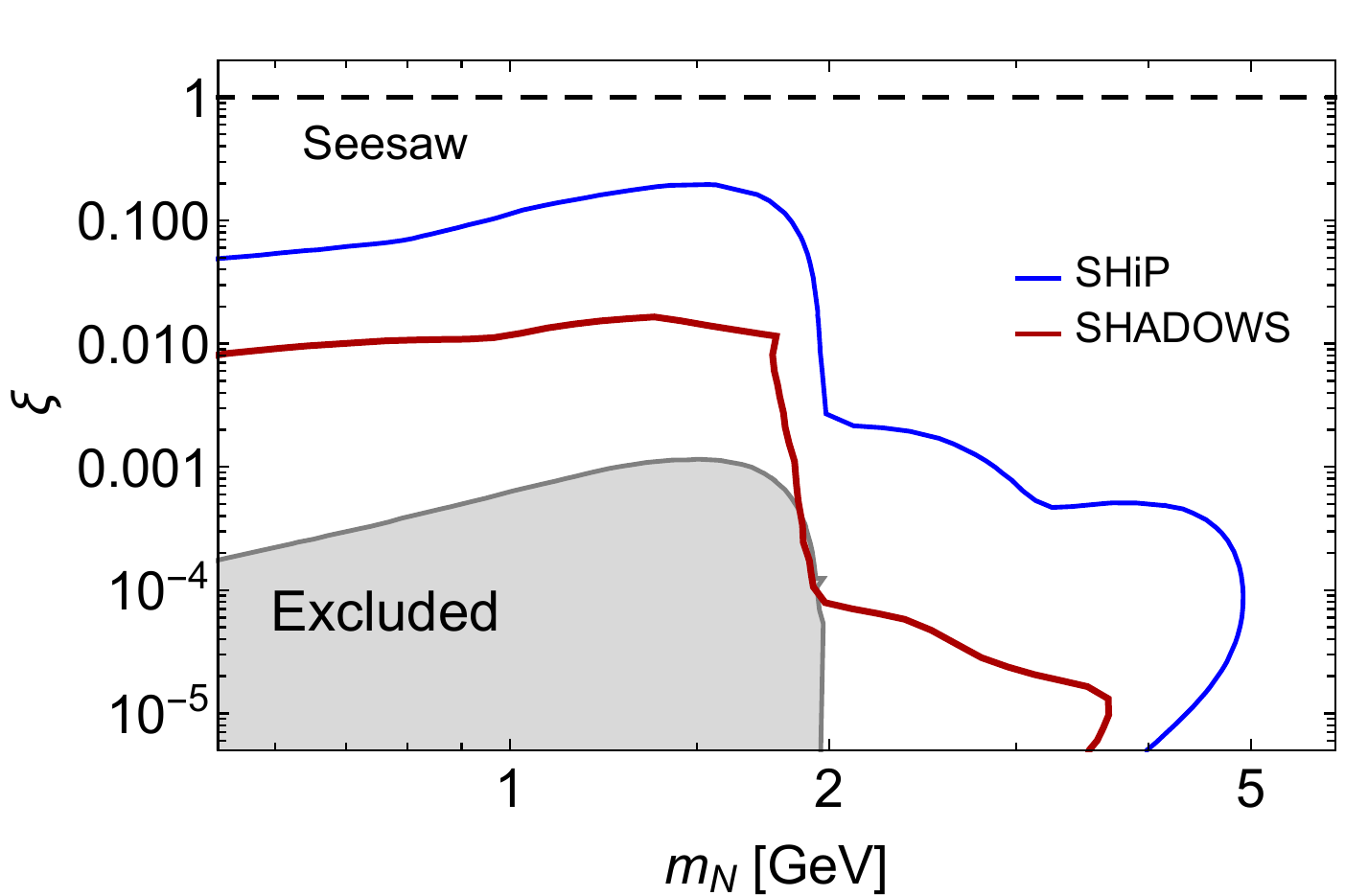}~\includegraphics[width=0.45\textwidth]{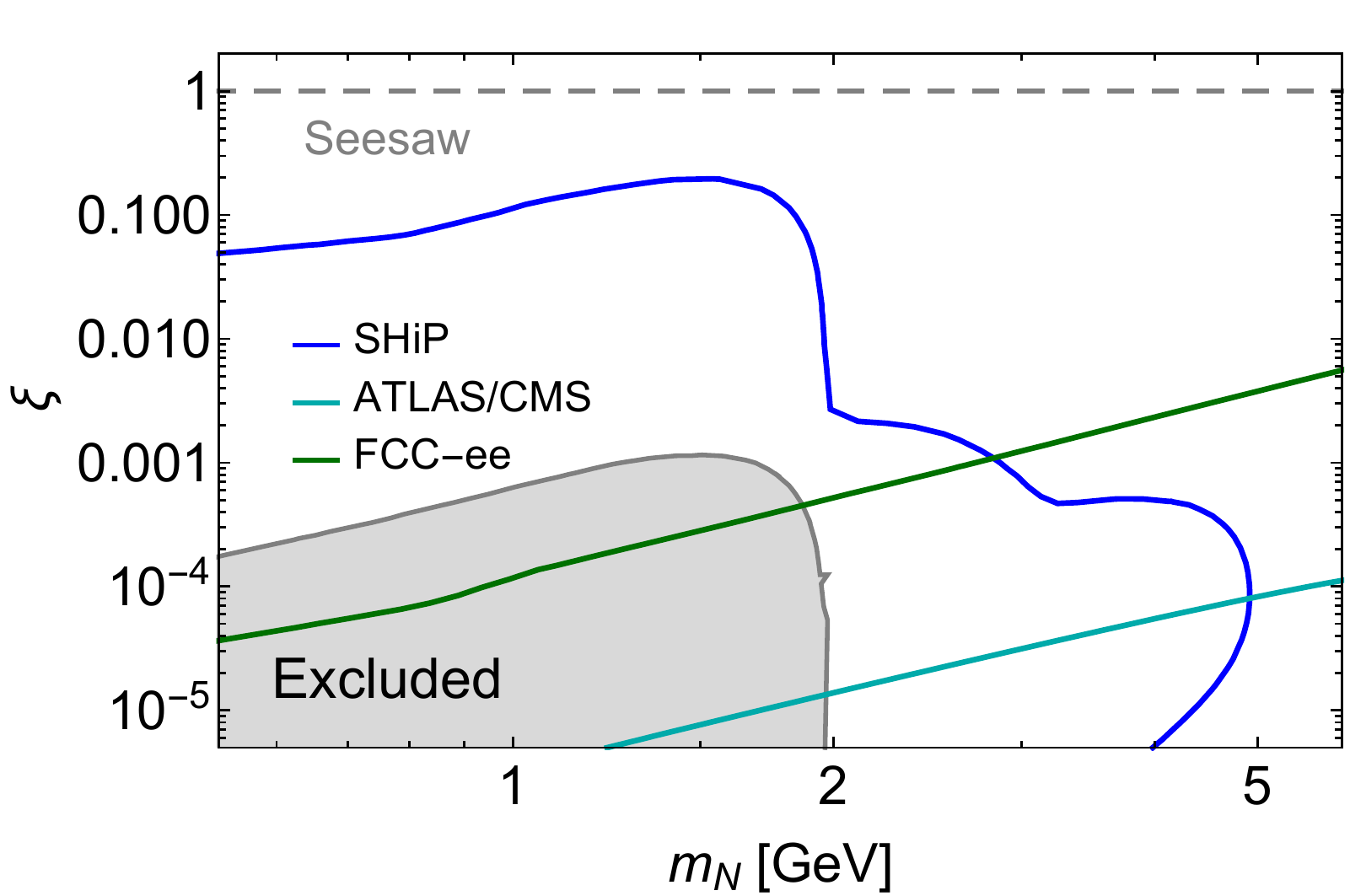}
    \caption{\textbf{The left panel}: the HNL parameter space reach of SHiP and SHADOWS in the plane $\xi = U^{2}_{\text{seesaw}}/U^{2}$-HNL mass. The departure of $\xi$ from the seesaw line $\xi = 1$ shows the scale of fine-tuning needed to explain neutrino masses by two HNLs with large couplings. \textbf{The right panel:} the same figure but with the reach of colliders included.}
    \label{fig:xi}
\end{figure*}

For SHiP, we also include the iso-contours corresponding to $N_{\text{events}}=100$ in Fig.~\ref{fig:summary}. Such a large number of events allows not only to establish the existence of a new particle but also identify its properties, such as branching ratios of various decay channels, precise mass, etc. We conclude that the combination of lower beam intensity and non-optimal geometric placement for HIKE and SHADOWS worsens their potential to explore the ``coupling frontier'' compared to SHiP. In particular, assuming that all the experiments operate in background-free regime, in the domain of small couplings, where SHADOWS and HIKE may detect 1 event at 90\% CL, SHiP would be able to reconstruct their parameters, such as mass, spin, and the probability of various decays. 

To stress the importance of the maximization of the FIP event yield, let us look closer at the model of HNLs. The main motivation for the HNLs is that they provide a very simple explanation of the observed active neutrino masses, in a way very similar to other fermions of the SM, i.e., by the mixing of left-handed and right-handed states via a Dirac mass term. For this, however,
(i) at least two HNLs are required in order to explain to observed neutrino mass differences $\Delta m_{\text{solar}}^{2},\Delta m_{\text{atm}}^{2}$;
(ii) their coupling constants should be at the level presented in Fig.~\ref{fig:light-FIPs-colliders} as the ``sea-saw line". As we see, such couplings are very far from what will be accessed by the current experiments.  It is, however, possible that the couplings of each of the HNLs are orders of magnitude larger, but that their contributions to the active neutrino masses cancel with high precision, due to a fine-tuning or a symmetry. The fine-tuning scale associated with this cancellation is $\xi = U^{2}_{\text{seesaw}}/U^{2}$, where $U^{2}_{\text{seesaw}}\sim 5\cdot 10^{-11}(1\text{ GeV}/m_{N})$ is the see-saw line which does not require the fine-tuning. As shown in Fig.~\ref{fig:xi}, beam dump experiments allow probing the part of the parameter space that is unreachable to collider experiments. In addition, we find that the layout of the SHiP experiment at the SPS ECN3 beam facility is close to optimal, assuming that the background is negligibly small.

\section*{Acknowledgements}
AB is supported by the European Research Council (ERC) Advanced Grant ``NuBSM'' (694896). KB is partly funded by the INFN PD51 INDARK grant. OM is supported by the NWO Physics Vrij Programme “The Hidden Universe of Weakly Interacting Particles” with project number 680.92.18.03 (NWO Vrije Programma), which is (partly) financed by the Dutch Research Council (NWO). MO received support from the European Union’s Horizon 2020 research and innovation program under the Marie Sklodowska-Curie grant agreement No. 860881-HIDDeN.

\newpage

\bibliographystyle{utphys}
\bibliography{bib.bib}

\newpage 
\appendix 

\section{Technical details of the sensitivity estimates}
\label{app:numerics}

\subsection{The number of events}

The expression for the number of events that we use looks as follows: 
    \begin{multline}
    N_{\text{events}} = \sum_{\text{mother}}N_{X,\text{prod}}\times \int dE_{X} \int \limits_{\text{decay volume}}d\theta_{X} dz_{X} \times \\ \times f_{X}(\theta_{X},E_{X})\times \epsilon_{\text{az}}(\theta_{X},z_{X}) \times \\ \times \frac{dP_{\text{dec}}}{dz_{X}} \times  \epsilon_{\text{dec}}(\theta_{X},E_{X},z_{X}) \times \text{Br}_{\text{vis}}(m_{X})
    \label{eq:Nevents}
\end{multline}
Here, $N_{X,\text{prod}} = N_{\text{mother}}\times P_{\text{mother}\to X}$ is the total number of particles $X$ produced by a mother particle, with $P_{\text{mother}\to X}$ being the probability of production. $z_{X},\theta_{X},E_{X}$ are, correspondingly, the longitudinal position, polar angle, and energy of the decaying FIP of type $X$, $f_{X}(\theta_{X},E_{X})$ is the differential distribution of FIPs in the polar angle and energy, and $\epsilon_{\text{az}}(\theta_{X},z_{X})$ is the azimuthal acceptance: 
    \begin{multline}
    \epsilon_{\text{az}} = \frac{1}{2\pi}\text{min}
\big[ \Delta \phi_{\text{decay volume}}(\theta_{X},z_{X}), \\ \Delta \phi_{\text{end of spectrometer}}(\theta_{X}))\big]
    \end{multline}
i.e. fraction of azimuthal angle for which FIPs \textbf{1)} decays inside the decay volume and \textbf{2)} have a trajectory intersecting the whole spectrometer.  The differential decay probability $\frac{dP_{\text{dec}}}{dz_{X}}$ is: 
\begin{equation}
\frac{dP_{\text{dec}}}{dz_{X}} = \frac{\exp[-l(z,\theta)/l_{\text{X,decay}}]}{l_{\text{X,decay}}} \frac{dl(z_{X},\theta_{X})}{dz_{X}},
\end{equation}
with $l = z_{X}/\cos(\theta_{X})$ being the displacement. The decay product acceptance $\epsilon_{\text{dec}}(\theta_{X},E_{X},z_{X})$ is the fraction of FIPs within the azimuthal acceptance whose decay products point to the detector. Finally, $\text{Br}_{\text{vis}}(m_{X})$ is the branching ratio of the FIP decay into final states visible at the given experiment.

Further details may be found in the works~\cite{Boiarska:2021yho,Boyarsky:2022epg,Ovchynnikov:2022its}, where the approach has been used for various experiments located at different facilities. 
In addition, the publicly available code \texttt{SensCalc}, which allows computing sensitivities using this approach, will be provided in a paper~\cite{Ovchynnikov:2023cry} soon to appear.

\subsubsection{Lower bound of the sensitivity}
In the regime of long FIP lifetimes, the exponent in the expression for the decay probability is 
\begin{equation}
\exp[-l/l_{\text{X,decay}}]\approx 1,
\end{equation}
and the number of events becomes
    \begin{equation}
    N_{\text{events}} \approx N_{X,\text{prod}} \times \frac{l_{\text{fid}}}{c\tau_{X}\langle E_{X}^{-1}\rangle^{-1}/m_{X}}\times \epsilon,
    \label{eq:Nevents-approx-1}
    \end{equation}
    where $\epsilon$ is the overall acceptance:
    \begin{multline}
    \epsilon \equiv \frac{1}{ l_{\text{fid}}}\int dE_{X} \int \limits_{\text{decay volume}}d\theta_{X} dz_{X} \times f_{X}(\theta_{X},E_{X})\times \\ \times \frac{\langle E^{-1}_{X}\rangle^{-1}}{E_{X}} \times\epsilon_{\text{dec}}(\theta_{X},E_{X},z_{X})\times \epsilon_{\text{az}}(\theta_{X},z_{X})
    \label{eq:acceptance}
\end{multline}

\subsection{Sensitivity to dark scalars}

In Fig.~\ref{fig:summary}, we report the 90\% CL sensitivity of SHADOWS to dark scalars. The sensitivity has been obtained assuming the setup described in the SHADOWS LoI~\cite{Alviggi:2839484} and requiring $N_{\text{events}}>2.3$. 

Let us discuss the setup in detail. The distance of the decay volume from the target is considered to be $l_{\text{min}} = 10\text{ m}$\footnote{Note that $l_{\text{min}}$ reported in the LoI in the section about experiment specifications is different, being 14\,m. The reason for this discrepancy is unclear to us.}, and the decay volume is defined to be located entirely off-axis with the closest decay volume wall being parallel to the beam-axis at a distance of 1\,m. We consider the given box geometry of the decay volume with length 20\,m and the cross-section equal to $2.5\times 2.5\text{ m}^{2}$, and use the detector cross-section equal to $2.5\times 2.5\text{ m}^{2}$. In addition, we require at least two charged decay products that point to the beginning of the dipole magnet, located $\simeq 2.5$ m downward the end of the decay volume. We note that the last specification may significantly overestimate the sensitivity of SHADOWS as the LoI specifies that the spectrometer and its downstream particle identification system, consisting of a calorimeter and muon system, have a total length of 12-14\,m. For proper identification of the decay products and reconstruction of their kinematics, their trajectories have to be contained within the acceptance of the entire detector, also requiring that the spectrometer magnet field is accounted for.  

As for the scalar phenomenology, we follow~\cite{Boiarska:2019jym}. For the probability of producing scalars, we use 
\begin{multline}
    \text{P}(B\to S) \approx \sum_{B_{x} = B^{0,+}}f_{b\to B_{x}}\times \\ \times \text{max}\left[\text{Br}_{\text{incl}}(B\to S), \text{Br}_{\text{excl}}(B\to S+X_{d/s})\right]
    \label{eq:br-ratio}
\end{multline}
Here, $f_{b\to B_{x}}$ is the fragmentation fraction ($f_{b\to B^{+}}\approx f_{b\to B^{0}}\approx 0.411$). $\text{Br}_{\text{incl}}(B\to S)$ is the inclusive branching ratio, which is equal to $\approx 6.6$ in the domain $m_{S}\ll m_{b}$. $\text{Br}_{\text{excl}}$ is the sum over all exclusive channels considered in~\cite{Boiarska:2019jym}, and $X_{d/s}$ is the final state appearing from the hadronization of $d$ or $s$ quarks. The inclusive description breaks down at large scalar masses~\cite{Boiarska:2019jym}; in addition, PBC recommended using the exclusive description~\cite{Beacham:2019nyx}. Nevertheless, in~\cite{Alviggi:2839484}, a purely inclusive estimate has been used. As a result, the branching ratio~\eqref{eq:br-ratio} matches the value used in~\cite{Alviggi:2839484} for scalars with masses $m_{S}\lesssim 2\text{ GeV}$, but is larger for larger masses.

For the distribution of mother particles, $B$ mesons, we use the angle-energy distribution and the production probability $\chi_{b\bar{b}} = 2.7\cdot 10^{-7}$ from~\cite{CERN-SHiP-NOTE-2015-009}, which also includes the cascade enhancement from the amount of $B$s produced in secondary interactions in the thick target.

Finally, we assume that the detectable decays of scalars are those containing at least two ``visible'' particles (photons or charged particles). In this way, we allow not only fully reconstructable states (where the kinematics may be fully reconstructed) but also partially reconstructed decays, containing particles such as neutrinos among the other decay products. Examples of the latter are decays $S\to \tau \bar{\tau} \to \nu_{\tau}\bar{\nu}_{\tau}+\text{visible}$, $S\to D\bar{D}\to \nu + \bar{\nu}+\text{visible}$, which dominate the scalar decays above di-$\tau$ mass threshold. In contrast, in the LoI, only fully reconstructable states have been required. We do not require any other cut, such as the minimum opening angle or the energy cuts. 

Therefore, given the discussion above, our sensitivity estimate must match the sensitivity from the LoI in the mass range $m_{S}\lesssim 2\text{ GeV}$. In contrast, for larger masses, it should be more optimistic.

\begin{figure}[!h]
    \centering\includegraphics[width=0.45\textwidth]{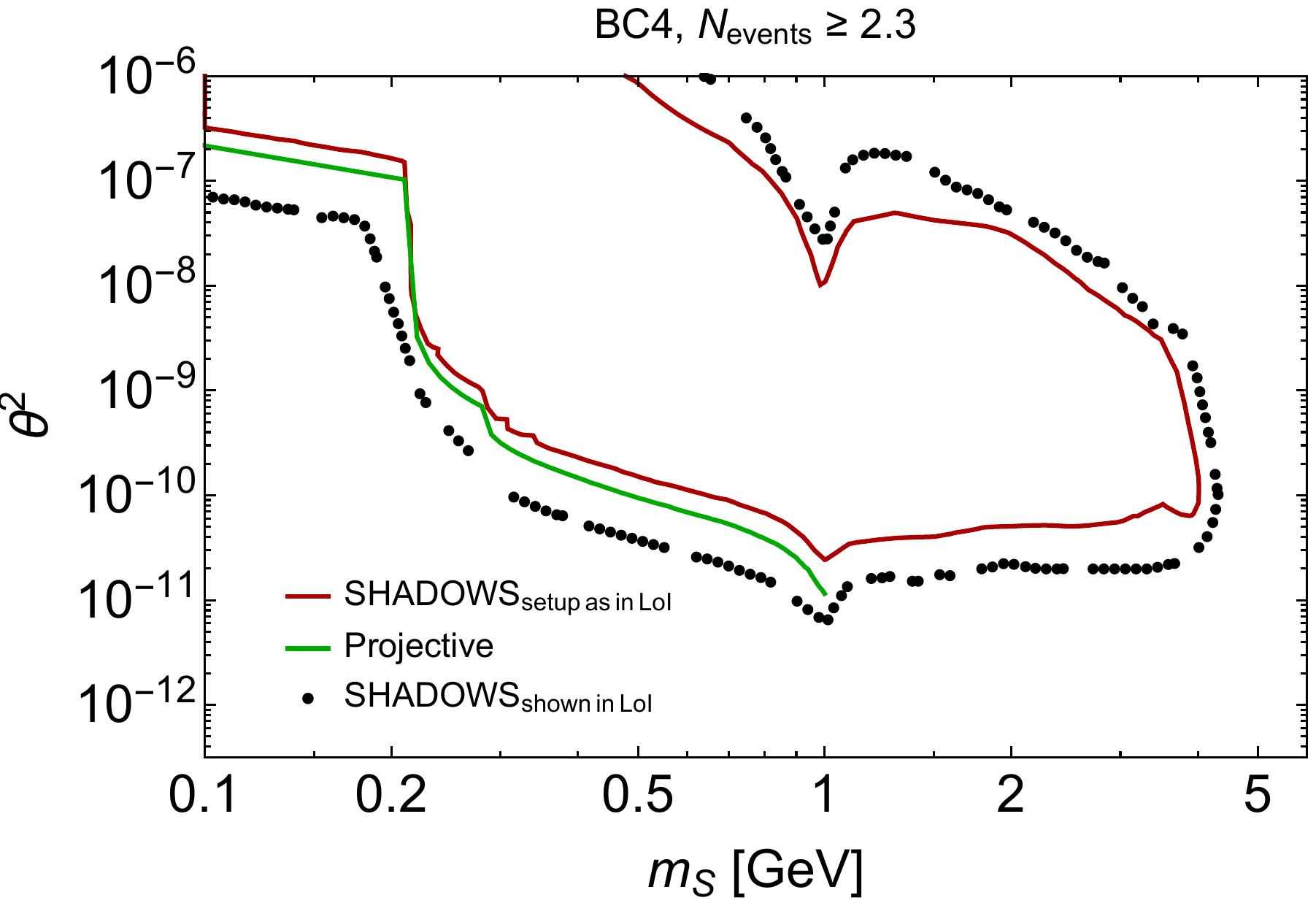}
    \caption{Sensitivity of SHADOWS to dark scalars with the mixing coupling. The red line shows the sensitivity obtained from our computation (Eq.~\eqref{eq:Nevents}) for the experimental setup used in the SHADOWS LoI~\cite{Alviggi:2839484}. The black dots show the sensitivity reported in the LoI. Finally, the green line shows the sensitivity of the projective setup where scalars are required to point to the SHADOWS detector acceptance window and with the decay products acceptance set to 1. The sensitivity of SHADOWS cannot be better than the sensitivity of the projective setup (see text for details).}
    \label{fig:SHADOWS-BC4}
\end{figure}

The comparison of the sensitivities of SHADOWS obtained by our calculations and by the calculations in the LoI is shown in Fig.~\ref{fig:SHADOWS-BC4}. 

We notice a significant difference at the lower and upper bounds of the sensitivity. Namely, at small masses, the number of events at the lower bound reported by the LoI is larger by up to a factor of 15 compared to our predictions. The situation is similar for large masses, where the sensitivity reported in the LoI should be much more conservative.

While we cannot identify the reason for such discrepancy, we can present a simple qualitative argument suggesting that the sensitivity presented in the LoI could be overestimated. Namely, let us consider an experiment with the projective decay volume along the lines pointing from the target to the SHADOWS detector window. For the latter, we choose the same plane with cross-section $2.5\times 2.5\text{ m}^{2}$ located 2.5 m downward the end of the decay volume.

It would not be possible to build such an experiment because of the limitations imposed by the infrastructure. However, what is important for us is that it must have better sensitivity than the real SHADOWS configuration. Indeed, its detector covers the same polar angles as the detector of SHADOWS, but, there would be no geometric shortening of the length of the decay volume for the scalars pointing to the detector, and there would be less azimuthal suppression.

\begin{figure}
    \centering
    \includegraphics[width=0.5\textwidth]{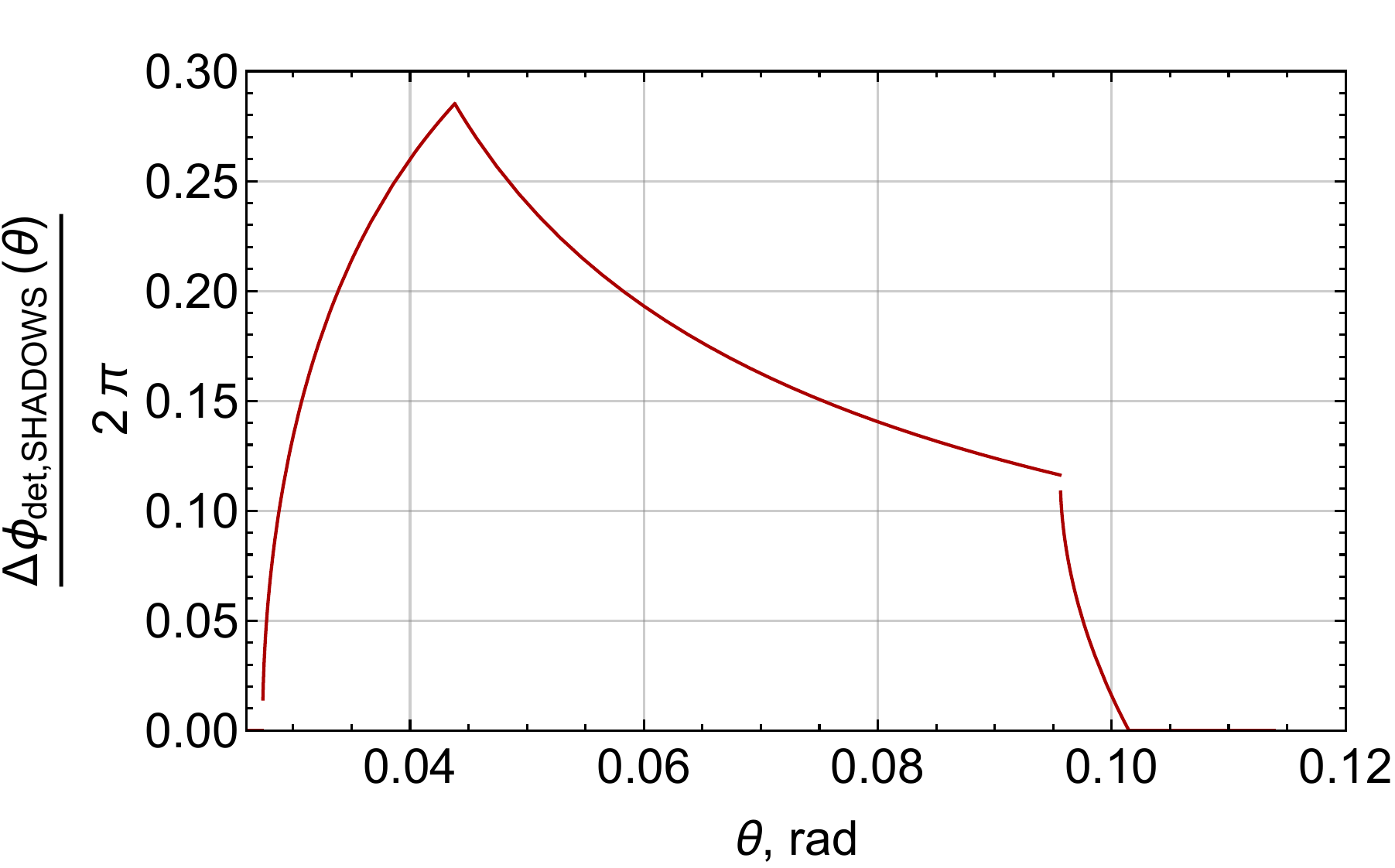}
    \caption{The fraction of the azimuthal angle covered by the SHADOWS detector window located 2.5 m downwards the end of the decay volume. The solid angle covered by the detector, $\Omega_{\text{det}} = \int \sin(\theta)\Delta \phi_{\text{det,SHADOWS}}d\theta$, matches the simple formula $S_{\text{det}}/L_{\text{to det}}^{2}\approx 4.6\cdot 10^{-3}\text{ sr}$.}
    \label{fig:shadows-detector-window}
\end{figure}

Further, let us consider the scalar mass $m_{S} \lesssim 1 \text{ GeV}$, for which the decay products fly roughly in the same direction as the decaying scalar, and we may approximate the decay products acceptance by its upper bound $\epsilon_{\text{dec}}\approx 1$. Finally, we may consider the lower bound of the sensitivity only. Under these assumptions, the number of decays is approximately
\begin{multline}
N_{\text{events}} \approx N_{\text{PoT}}\times \chi_{b\bar{b}}\times 2(f_{b\to B^{+}}+f_{b\to B^{0}})\times \text{Br}_{B
\to S}\times  \\ \times \epsilon_{\text{geom}}\times\frac{l_{\text{fid}}\cdot m_{S}\langle p^{-1}_S\rangle}{c\tau_{S}},
\label{eq:simple-estimate-scalars}
\end{multline}
where $\epsilon_{\text{geom}}$ is the fraction of the scalars pointing to the detector, and $\langle p^{-1}_S\rangle$ is the mean inverse momentum of such scalars. These quantities may be obtained just from knowing the scalar distribution function, which may be easily derived given the simple kinematics of the scalar production in 2-body decays $B\to S+X_{s}$, and the azimuthal coverage of the SHADOWS detector window, which is shown in Fig.~\ref{fig:shadows-detector-window}.

The values of all the relevant parameters for the particular point $m_{S} = 150\text{ MeV}$ are reported in Table~\ref{tab:point}.

\begin{table*}
\centering
    \begin{tabular}{|c|c|c|c|c|c|c|c|}
      \hline Parameter& $\text{Br}(B\to S)$& $\epsilon_{S}$& $m_{S}\langle p_{S}^{-1}\rangle$& $c\tau_{S}, \text{ m}$ & $N_{\text{events}}$ \\ \hline
    Value & $\approx 6.6$ & $0.09$ & $3.7\cdot 10^{-3}$ & $7.7\cdot 10^{-3}/\theta^{2}$ & $2.3\left( \frac{\theta^{2}}{1.4\cdot 10^{-7}}\right)^{2}$ \\ \hline
    \end{tabular}
    \caption{The values of the parameters entering Eq.~\eqref{eq:simple-estimate-scalars} for the particular scalar mass $m_{S}=150\text{ MeV}$.}
    \label{tab:point}
\end{table*}

Comparing the lower bound predicted by this formula with the sensitivity from the SHADOWS LoI, we find that the sensitivity of the projective experiment lies \textit{well above} the latter, see Fig.~\ref{fig:SHADOWS-BC4}. In contrast, it lies below our estimate. At small masses, it is very close to our estimate. The increasing difference at large masses is explained by the decay products acceptance which is assumed to be 1 for the projective setup.

\end{document}